\documentclass[conf]{new-aiaa}
\usepackage[utf8]{inputenc}

\usepackage[export]{adjustbox}
\usepackage{graphicx}
\usepackage{floatrow}
\usepackage{subfig}
\usepackage{caption}
\usepackage{amsmath}
\usepackage[version=4]{mhchem}
\usepackage{siunitx}
\usepackage{longtable,tabularx}
\title{Stochastic Backscatter for Grey-Area Mitigation\\ in Hybrid RANS-LES Simulations}

\author{A. Passariello \footnote{Ph.D. student, Department of Industrial Engineering (\url{angelo.passariello@unina.it}).}}
\affil{University of Naples Federico II, Naples, 80125, Italy}
\author{P. Catalano \footnote{Senior Research Engineer, Fluid Dynamic Modelling Group (\url{p.catalano@cira.it}).} and C. De Lucia \footnote{Researcher, High-Performance Computing Group (\url{c.delucia@cira.it})}}
\affil{Italian Aerospace Research Centre, Capua, Caserta, 81043, Italy}
\author{R. Tognaccini \footnote{Full professor in Aerodynamics, Department of Industrial Engineering (\url{rtogna@unina.it}).}}
\affil{University of Naples Federico II, Naples, 80125, Italy}

\begin{document}

\maketitle

\begin{abstract}
A scale-resolving simulation methodology that includes stochastic energy backscatter is incorporated into a proprietary block-structured compressible flow solver. 
Particular attention is devoted to the discretisation of the convective terms in the averaged/filtered governing equations. The objective is to achieve satisfactory dissipation and dispersion properties, while minimising the number of modifications to be made to the existing RANS solver, which employs, by default, a second-order accurate central scheme with Jameson-Schmidt-Turkel scalar artificial dissipation.
A novel blending strategy, combining non-dissipative and strongly dissipative numerical discretisations, is proposed to enhance the overall numerical stability. 
First of all, the model is calibrated using the classic decay of isotropic homogeneous turbulence. Then, its effectiveness in mitigating the grey area is shown through the simulation of the mixing co-flow between the wake originating downstream of an airfoil at zero angle of attack and the zero-pressure-gradient turbulent boundary layer developing over a flat plate.  
\end{abstract}

\section*{Nomenclature}

{\renewcommand\arraystretch{1.0}
\noindent\begin{longtable*}{@{}l @{\quad=\quad} l@{}}
$Re$  & Reynolds number \\
$k$   & turbulent/subgrid kinetic energy \\
$\epsilon_k$ & turbulent/subgrid kinetic energy dissipation rate \\
$\omega$ & turbulence frequency \\
$\nu_t$ & eddy/subgrid viscosity \\
$\Delta$ & LES filter width \\
$\tau_{\scalebox{0.5}{SGS}}$ & subgrid timescale \\
$\nu$ & dynamic viscosity \\
$\kappa_\nu$ & Von Karman constant ($=0.41$) \\
$d_w$ & wall distance \\
$\boldsymbol{x}$ & position vector \\
$\rho$ & averaged/filtered density \\
$\boldsymbol{u}$ & averaged/filtered velocity \\
$\boldsymbol{u'}$ & fluctuating velocity \\
$V_\infty$ & freestream velocity \\
$u,v,w$ & velocity components in physical space \\
$\tau_{ij}$ & modelled stress tensor \\
$f_d$ & shielding function \\
$\Tilde{f}_d$ & modified shielding function \\
$\mathcal{N}(\mu_N,\sigma_N^2)$ & Gaussian probability distribution of mean $\mu_N$ and variance $\sigma_N^2$ \\
$\boldsymbol{I}$ & identity tensor \\
$c$ & local sound speed \\
$\gamma$ & specific heat ratio \\
$p$ & averaged/filtered static pressure \\
$\mathrm{Var}\{X\}$ & variance of the stochastic variable $X$ \\
$\mathrm{Cov}\{X,Y\}$ & covariance of the stochastic variables $X$ and $Y$

\end{longtable*}}

\section{Introduction}
\label{sec:intro}

\lettrine{N}{owadays}, Reynolds-Averaged Navier-Stokes (RANS) simulations are routinely employed for the preliminary design of novel air vehicles. 
By time-averaging the Navier-Stokes equations, RANS approaches completely avoid resolving small-scale turbulent structures, thus achieving a practical balance between accuracy and computational cost, especially for high-Reynolds-number flows. However, the predictive capability of RANS simulations rapidly deteriorates in the presence of massive separation and aerodynamic interference effects, where the unsteady nature of turbulence is evident.
These conditions are typically encountered at the margins of aircraft flight envelopes (e.g., in take-off and landing configurations \cite{goc2021large}) due to the combination of relatively high angles of attack and the deployment of auxiliary lifting surfaces (i.e., flaps and/or slats). 
Such complex phenomena cannot be modelled as is done in low-fidelity (RANS-based) solution approaches. 
This limitation justifies the increasing popularity of Scale-Resolving Simulation (SRS) methods, which, as suggested by their name, resolve at least part of the turbulence spectrum.
Among the aforementioned methodologies, a prominent example for industrial applications is Large Eddy Simulation (LES), which is based on a spatially filtered version of the Navier-Stokes equations. By means of varying formulations \cite{arvidson2017methodologies}, the LES filter width is dependent on the local mesh size, so that the resolving capability of LES is directly related to the grid employed.
Choi \& Moin \cite{choi2012grid} proved that the total number of grid points required to perform a (full) LES of a wall-bounded flow scales with $Re^{13/7}$. Although this scaling is more favourable than the one for a Direct Numerical Simulation (DNS), namely $Re^{37/14}$, the quasi-quadratic dependence on the Reynolds number makes such an approach unfeasible for industrially relevant configurations at present \cite{bose2018wall}. 

Since the mathematical structure of RANS and LES governing equations is basically identical \cite{heinz2020review} (if the commutation errors in the filtered Navier-Stokes equations are neglected), a variety of
approaches for combining the two simulation methodologies have been suggested so far: Wall-Modelled LES (WMLES) \cite{bose2018wall}, Detached
Eddy Simulation (DES) \cite{spalart2009detached},
Reynolds-Stress-Constrained LES (RSC-LES) \cite{chen2012reynolds}, UNIfied RANS-LES
(UNI-LES) \cite{heinz2007unified}, Partially Averaged Navier-Stokes
(PANS) \cite{girimaji2003pans}, Partially Integrated Transport Modeling (PITM) \cite{chaouat2005new}, and Scale Adaptive Simulation (SAS) \cite{menter2003scale} are some well-known representatives. The previous list is not meant to be exhaustive and the reader is referred to \cite{heinz2020review,menter2021overview} for a more general discussion about Hybrid RANS-LES Methods (HRLMs).

Among the approaches mentioned above, the one that undoubtedly stands out for its simplicity and ease of incorporation into existing finite-volume RANS-based flow solvers is DES. This method consists in using RANS to model the flow in close proximity to solid surfaces, where small-scale turbulence is predominant, and gradually switching to LES away from walls to resolve larger energy-containing eddies.
The switch is dynamically and locally determined based on the comparison between the characteristic length scales implied by the two approaches; the true transition is then accomplished by substituting the minimum length scale in the underlying turbulence model. 
Since they can automatically adapt to different flow resolutions, non-zonal DES-like methods are marked out by great flexibility. This makes them suitable for the simulation of flows characterised by a mixture of attached boundary layers and separated shear layers.
However, several problems persist in the basic formulation of DES. For example, it was observed that, if the DES limiter is activated inside boundary layers, the consequent reduction of eddy/subgrid viscosity may result in Modelled-Stress Depletion (MSD) and Grid-Induced Separation (GIS) \cite{menter2004adaptation}. 
A viable solution to this undesired behaviour is to "shield" the boundary layer from the limiter action by means of an appropriate shielding function \cite{spalart2006new}. The resulting class of methods is given the name of Delayed DES (DDES).

Another issue which is frequently encountered when dealing with DES is the spatial delay in the formation of turbulent fluctuations after the switch from RANS mode to LES mode. This detrimental phenomenon, commonly referred to as Grey Area (GA) \cite{arvidson2017methodologies}, is due to the fact that the relatively high eddy viscosity coming from the RANS flow region is partially advected into the LES region, where it damps the development of resolved turbulent stresses.
Obviously, the GA phenomenon is much more frequently encountered in the simulation of turbulent flows where low-frequency unsteadiness is almost negligible (e.g., attached boundary layers or mixing layers), rather than in the simulation of massively separated flows, where, indeed, natural fluctuations trigger the formation of resolved turbulence.

GA mitigation is the main topic of the present paper. In the literature on HRLMs, two different approaches are typically pursued: reducing the subgrid stresses in the initial part of shear layers (e.g., by modifying the definition of the LES filter width \cite{arvidson2017methodologies}) and/or adding some form of stochastic forcing to the momentum equations \cite{kok2010destabilizing,kok2017stochastic}.
Here, the focus is on the second class of the aforementioned methodologies. 
In particular, the Stochastic BackScatter (SBS) model conceived by Kok \cite{kok2017stochastic} has sparked interest in the authors for the promising results recently obtained from the CFD simulation of a delta wing at high angle of attack and of a three-element airfoil at low incidence \cite{kok2018application}, both showing nearly no GA in separating shear layers.

The need to incorporate the SBS model into the proprietary CFD code developed at the Italian Aerospace Research Centre (CIRA), i.e., the \underline{U}nsteady \underline{Z}onal \underline{E}uler \underline{N}avier-Stokes flow solver (UZEN), while limiting the impact of the modifications to the solver itself, brought about a number of challenges to be faced. The adopted solutions have introduced several novelties in the algorithm which, in our opinion, are worth being presented here. 
For the discretisation of the convective terms in the governing equation, CIRA's flow solver employs, by default, a second-order accurate central scheme with Jameson-Schmidt-Turkel (JST) scalar artificial dissipation. 
The latter, as will be shown in Section \ref{subsec:calib}, is not appropriate for SRSs due to its highly dissipative nature.
On the other hand, the implementation of the fourth-order low-dispersion symmetry-preserving finite-volume scheme proposed by Kok \cite{kok2009high} would have involved numerous complications (especially for computations) due to the need for an extended stencil.
The optimal compromise has been found in the second-order Low-Dissipation Low-Dispersion (LD2) scheme by L\"owe et al. \cite{lowe2016low}.
To be precise, a blend of the LD2 scheme and a first-order upwind scheme is proposed to enhance the overall stability of the numerical method. 
This hybrid discretisation is employed only in the X-LES blocks selected by the user, whereas the original JST scheme is still used in RANS flow regions.

The objective of this dissertation is to prove the feasibility of incorporating the SBS model into existing second-order accurate RANS-based flow solvers and to further test its performance in the simulation of complex flow cases of aeronautical interest.
The paper is organised as follows: in Section \ref{sec:model} the formulation of the SBS model (in the eXtra-Large Eddy Simulation framework \cite{kok2004extra}) is briefly summarised; in Section \ref{sec:numerics} details on the numerical method are provided; in Section \ref{sec:results} the model is firstly calibrated on the classic decay of isotropic homogeneous turbulence and, finally, its effectiveness in the GA mitigation is shown for a test case involving wake-boundary-layer mixing.

\section{Model description}
\label{sec:model}

\subsection{eXtra-Large Eddy Simulation (X-LES)}
The scale-resolving approach employed in UZEN flow solver is the eXtra-Large Eddy Simulation (X-LES). The latter is a non-zonal DES-type hybrid methodology consisting in the blend of
the TNT $k-\omega$ turbulence model \cite{kok2000resolving} for the RANS mode and the $k-$equation subgrid model for the LES mode \cite{kok2004extra}.
The simulation branch is determined locally and dynamically by taking the minimum between the RANS and the LES length scales ($\ell_{\scalebox{0.5}{RANS}}$ and $\ell_{\scalebox{0.5}{LES}}$, respectively): 

\begin{equation}
    \ell = \min(\ell_{\scalebox{0.5}{RANS}},\ell_{\scalebox{0.5}{LES}}) = \min\left(\sqrt{k}/\omega,C_1\Delta\right)
    \label{eq:turlen}
\end{equation}

\noindent
The model constant $C_1$ results from the calibration of the method on the classic test case of Decaying Isotropic Homogeneous Turbulence (DIHT, see Section \ref{subsec:calib} for details). The LES filter width $\Delta$ is set equal to the maximum cell size in the three coordinate directions of the computational space, according to the original definition proposed by Spalart for DES97 \cite{spalart2009detached}.
In practice, the switch from RANS to LES - and vice versa - is accomplished by substituting the blended length scale into the definitions of both the eddy/subgrid viscosity and the turbulent/subgrid kinetic energy dissipation rate: 

\begin{equation}
    \nu_t = \sqrt{k}\,\ell,
    \quad
    \epsilon_k = \beta\,{k^{3/2}}/{\ell}
\end{equation}

\noindent
with $\beta=0.09$. To avoid the  of GIS, the shielding function, defined in context of DDES \cite{spalart2006new}, is also included in the formulation:

\begin{equation}
    f_d = 1-\tanh\left[(8r_d)^3\right],
    \quad
    r_d = \frac{\nu_t+\nu}{\sqrt{\frac{\partial u_i}{\partial x_j}\frac{\partial u_i}{\partial x_j}}\,\kappa_\nu^2\,d_w^2}
    \label{eq:shielding}
\end{equation}

\noindent
The computation of $f_d$ requires the preliminary evaluation of the wall distance field $d_w$. To this aim, an efficient algorithm is proposed in Section \ref{subsec:wall_dist}. Thanks to the introduction of $f_d$, Eq. \eqref{eq:turlen} can be modified as follows:

\begin{equation}
    \ell = (1-f_d)\ell_{\scalebox{0.5}{RANS}} + f_d\,\min(\ell_{\scalebox{0.5}{RANS}},\ell_{\scalebox{0.5}{LES}})
\end{equation}

\noindent
Since the shielding function goes to zero in close proximity to solid walls (where $d_w$ is relatively small), $\ell$ practically coincides with $\ell_{\scalebox{0.5}{RANS}}$ in attached boundary layers, thus avoiding the erroneous activation of the LES mode in the case of ambiguous grid densities \cite{spalart2006new}. For the sake of consistency with the literature, this variant of the method is given the name of Delayed X-LES.

An additional enhancement to the simulation methodology, aiming to mitigate the GA issue, consists in filtering the velocity field in time before computing the subgrid stresses \cite{kok2017stochastic}. This has the effect of reducing the modelled part of the turbulence spectrum in free shear layers, which are characterised by relatively high mean velocity gradients. As a consequence, the earlier development of resolved turbulence is facilitated. In mathematical terms: 

\begin{equation}
    u'_i(\boldsymbol{x},t) = u_i(\boldsymbol{x},t)-\dfrac{\Tilde{f}_d}{t}\int_0^t u_i(\boldsymbol{x},t^*)\,dt^*,
    \quad
    \Tilde{f}_d =
    \begin{cases}
        f_d\quad &\text{if}\;\ell_{\scalebox{0.5}{RANS}}>\ell_{\scalebox{0.5}{LES}}\\
        0\quad &\text{if}\; \ell_{\scalebox{0.5}{RANS}}\leq\ell_{\scalebox{0.5}{LES}}
    \end{cases}
    \label{eq:HPF}
\end{equation}

\begin{equation}
    \dfrac{\tau_{ij}}{\rho} = \nu_t\left(\dfrac{\partial u'_i}{\partial x_j}+\dfrac{\partial u'_j}{\partial x_i}-\dfrac{2}{3}\dfrac{\partial u'_m}{\partial x_m}\delta_{ij}\right) - \dfrac{2}{3} k\delta_{ij}
\end{equation}

\noindent
In Eq. \eqref{eq:HPF}, $\Tilde{f}_d$ simply represents a mathematical artifice to identify the local simulation mode. Since it is identically zero in RANS mode, the High-Pass Filter (HPF) is effectively switched off, thus allowing the Reynolds stress tensor to recover its classic definition. Including $\Tilde{f}_d$, the blended characteristic length scale can finally be reformulated as follows:

\begin{equation}
    \ell = (1-\Tilde{f}_d)\,\ell_{\scalebox{0.5}{RANS}} + \Tilde{f}_d\,\ell_{\scalebox{0.5}{LES}}
\end{equation} 

\subsection{Stochastic BackScatter (SBS)}
\label{subsec:SBS}

The second strategy for mitigating the GA phenomenon relies on modelling the energy backscatter from the subgrid scales to the resolved scales. To this aim, the definition of the modelled stress tensor is modified as follows:

\begin{equation}
    \dfrac{\tau_{ij}}{\rho} = \nu_t\left(\dfrac{\partial u'_i}{\partial x_j}+\dfrac{\partial u'_j}{\partial x_i}-\dfrac{2}{3}\dfrac{\partial u'_m}{\partial x_m}\delta_{ij}\right) - \dfrac{2}{3} k\delta_{ij} - \Tilde{f}_d \,R_{ij}
    \label{eq:tau_def}
\end{equation}

\noindent
where $R_{ij}$ is a stochastic tensor. The divergence of $\boldsymbol{R}$ is modelled as the curl of a stochastic potential vector, $\boldsymbol{\xi}$:

\begin{equation}
    \boldsymbol{\nabla}\cdot\boldsymbol{R} = \boldsymbol{\nabla}\,\times\,(C_B\,k\,\boldsymbol{\xi})
    \label{eq:R_def}
\end{equation}

\noindent
The constant $C_B$ is set to one by default.
It can be demonstrated that the functional form of Eq. \eqref{eq:R_def} is the only one that ensures a backscatter rate consistent with the turbulence theory \cite{kok2017stochastic}.
After a bit of algebra, the analytical expression of $\boldsymbol R$ can be obtained:

\begin{equation}
    R_{ij} = \varepsilon_{jim}\,C_B\,k\xi_m
    \label{eq:R_expr}
\end{equation}

\vspace{2pt}
\noindent
where $\varepsilon$ denotes the Levi-Civita symbol. For the sake of clarity, Eq. \eqref{eq:R_expr} is rewritten below in matrix notation:
\vspace{2pt}

\begin{equation}
    \boldsymbol{\boldsymbol{R}} = C_B\,k\:
    \begin{bmatrix}
        0 & -\xi_3 & \xi_2 \\
        \xi_3 & 0 & -\xi_1 \\
        -\xi_2 & \xi_1 & 0
    \end{bmatrix}
\end{equation}

\newenvironment{nobreaks}{\vbox\bgroup}{\egroup}
\vfill
\noindent
For a detailed explanation of the rationale behind this particular definition of the stochastic tensor, the reader is referred to \cite{kok2017stochastic}.
Here, only the implications of Eq. \eqref{eq:R_def} are briefly summarised:
\vspace{5pt}

\begin{itemize}
    \item Firstly, the magnitude of the backscatter term depends on the subgrid kinetic energy. A high value of $k$ is likely to be convected into the initial region of free shear layers when the upstream flow is treated in RANS mode. Consequently, the stochastic forcing becomes locally quite intense, leading to the formation of resolved turbulence. Moving downstream, the modelled portion of the turbulence spectrum gradually diminishes, as does $k$. As a result, the random forcing weakens. In particular, when the turbulent flow structures are fully resolved, the backscatter term vanishes, ensuring physical consistency with DNS.
    \vspace{5pt}
    \item Secondly, the presence of $\Tilde{f}_d$ in Eq. \eqref{eq:tau_def} ensures that the random source term in the momentum equations is set to zero in RANS mode and that it is smoothly switched on at increasing distances from solid walls.
    \item Finally, as evident from Eq. \eqref{eq:R_def}, the stochastic source term is solenoidal. Therefore, according to Lighthill's acoustic analogy, it cannot generate noise. This makes the method suitable for aeroacoustic simulations.
\end{itemize}
\vspace{5pt}

The properties of the stochastic vector $\boldsymbol{\xi}$ must still be specified. Its components are required to follow a standard normal distribution, i.e., a Gaussian probability density function with zero mean and unit variance. Furthermore, each component of $\boldsymbol\xi$ is constructed to be spatially uncorrelated over distances exceeding the LES filter width and temporally uncorrelated over intervals longer than the subgrid time scale ($\tau_{\scalebox{0.5}{SGS}}  \propto \Delta/\sqrt{k}$).
However, especially for the strongly anisotropic grids typically employed to resolve free shear layers, $\Delta$ may be significantly greater than the distance between adjacent grid points. In such cases, a certain degree of spatial correlation must be maintained for distances smaller than $\Delta$.
A similar argument applies to temporal correlations, which must be defined for time steps smaller than the subgrid time scale, $\tau_{\scalebox{0.5}{SGS}}$. 
The most appropriate mathematical form for these correlations is an exponential decay:

\begin{equation}
    \left\langle \xi_i(\boldsymbol{x},t) \,\xi_j(\boldsymbol{y},s)\right\rangle = \delta_{ij}\,\exp\left({-\frac{|\boldsymbol{x}-\boldsymbol{y}|^2}{2\,C_\Delta \Delta^2}}\right)\,\exp\left({-\frac{|t-s|}{\tau_{\scalebox{0.5}{SGS}}}}\right),
    \quad
    \tau = C_\tau \dfrac{\Delta}{\sqrt{k}}
    \label{eq:csi_corr}
\end{equation}

\noindent
where $C_\Delta = 0.1$ and $C_\tau = 0.05$. To ensure that Eq. \eqref{eq:csi_corr} is satisfied, a stochastic differential equation for the variable $\boldsymbol{\xi}$ must be solved \cite{kok2017stochastic}. The procedure involves three steps:
\vspace{5pt}
\begin{enumerate}
    \item Drawing values for a normally-distributed stochastic differential, say $dV_i(\boldsymbol{x},t)$, which must be both spatially and temporally uncorrelated:
    \vspace{2pt}
    \begin{equation}
        dV_i(\boldsymbol{x},t) = \mathcal{N}(0,d\boldsymbol{x}\,dt)
        \label{eq:dVi}
    \end{equation}
    \item Performing an implicit smoothing of $dV_i$ to obtain the spatially-correlated stochastic differential $dW_i$:
    \begin{equation}
        \left[\prod_{j=1}^3 dx_j\left(1-C_\Delta \Delta^2\,\dfrac{\partial}{\partial x_j^2}\right)\right]dW_i = 8 (\sqrt{C_\Delta}\Delta)^{3/2}\,dV_i
        \label{eq:smooth}
    \end{equation}
    \item Finally, solving a Langevin-type differential equation for $\rho \xi_i$ to add the desired exponentially-decaying temporal correlation:
    \vspace{2pt}
    \begin{equation}
        \rho\xi_i\,dt + \tau\left(\dfrac{\partial\,\rho\xi_i}{\partial t} + \dfrac{\partial\, \rho u_j \xi_i}{\partial x_j}\right) \,dt = \sqrt{2\tau}\,\rho\,dW_i 
        \label{eq:Langevin}
    \end{equation}
\end{enumerate}

\noindent
The numerical discretisation of Eqs. \eqref{eq:dVi}-\eqref{eq:Langevin} is discussed in Section \ref{subsec:stoc_eqs}.

\section{Numerical method}
\label{sec:numerics}

The SBS model has been incorporated into CIRA's in-house CFD code, UZEN. The solver employs a cell-centred finite-volume method on multi-block structured grids. Temporal integration is performed using a second-order accurate backward Euler scheme, combined with a dual-time stepping strategy. Inner iterations are advanced through an explicit low-storage multistage Runge–Kutta method.
Convergence acceleration in dual-time is optionally achieved using a Full-Approximation Storage (FAS) multigrid method, implicit residual averaging, and local time stepping.
For RANS simulations, the default spatial discretisation in UZEN relies on central differencing, complemented by Jameson-Schmidt-Turkel (JST) scalar artificial dissipation. However, the excessively dissipative nature of the latter scheme makes it unsuitable for SRSs (see Section \ref{subsec:calib}). 
In his work \cite{kok2017stochastic}, Kok employs a fourth-order low-dispersion symmetry-preserving finite-volume scheme \cite{kok2009high}, which ensures local conservation of kinetic energy by convection. Skew-symmetry preservation enhances stability and avoids the potential interference between numerical inaccuracy and modelling errors.
However, the fourth-order scheme makes use of a seven-point stencil in each computational direction. Incorporating it in UZEN would require significant effort because of the new treatment of the internal coupling between adjacent blocks.
Thus, an optimal balance between numerical accuracy and ease of implementation is to be sought. This compromise is achieved with the Low-Dissipation Low-Dispersion (LD2) second-order scheme proposed by Löwe et al. \cite{lowe2016low}, which extends the original three-point stencil employed for central differencing to a five-point stencil (by also using gradients when computing cell-face values). 
This approach is particularly convenient since Green-Gauss gradients are already computed elsewhere in the code and are readily available for use.
Therefore, the LD2 scheme has been incorporated into UZEN as part of the present work.

The remainder of this Section is organised as follows: details about the LD2 scheme, as well as a practical demonstration of its beneficial numerical properties, are presented in Section \ref{subsec:convec}. The algorithm used to compute the distance to the nearest wall, which is required for the evaluation of the shielding function, is described in Section \ref{subsec:wall_dist}. Finally, the numerical discretisation of the stochastic equations (previously introduced in Section \ref{subsec:SBS}) is discussed in Section \ref{subsec:stoc_eqs}.

\subsection{Convective fluxes}
\label{subsec:convec}

The convective terms in the averaged/filtered governing equations are discretised using the LD2 scheme proposed by Löwe et al. \cite{lowe2016low}. This is based on the skew-symmetric formulation introduced by Kok \cite{kok2009high}, which ensures both local and global conservation of kinetic energy, thus preventing the onset of numerical instabilities. In the LD2 formulation, the convective fluxes at cell-face centres are evaluated as follows:
\vfill
\begin{equation}
\boldsymbol{{F}}_{LR} =
\begin{bmatrix}
\dfrac{1}{2}(\rho_L \boldsymbol{u}_L + \rho_R \boldsymbol{u}_R)\\[5pt]
\dfrac{1}{2}(\rho_L \boldsymbol{u}_L + \rho_R \boldsymbol{u}_R)\,\dfrac{1}{2}(\boldsymbol{u}_L+\boldsymbol{u}_R) + \dfrac{1}{2}(p_L+p_R)\boldsymbol{I}\\[5pt]
\dfrac{1}{2}(\rho_L \boldsymbol{u}_L + \rho_R \boldsymbol{u}_R)\left[\dfrac{1}{2}(\boldsymbol{u}_L\cdot\boldsymbol{u}_R)+\dfrac{c_L\,c_R}{\gamma(\gamma-1)}\right] + \dfrac{1}{2}(\boldsymbol{u}_L p_R+\boldsymbol{u}_R p_L)
\end{bmatrix}
\label{eq:FLR}
\end{equation}

\noindent
The subscripts $L$ and $R$ respectively denote the "left" and "right" sides of the face shared by two adjacent control volumes in any direction of the computational space. The left and right values of velocity and pressure in Eq. \eqref{eq:FLR} are extrapolated from their respective cell-centre values (subscripts $\ell$ and $r$) through a local Taylor-series expansion:

\begin{equation}
\boldsymbol{u}_L = \boldsymbol{u}_\ell + \alpha\,(\boldsymbol{x}_r-\boldsymbol{x}_\ell)\cdot(\boldsymbol{\nabla}\boldsymbol{u})_\ell\;,
\quad
\boldsymbol{u}_R = \boldsymbol{u}_r - \alpha\,(\boldsymbol{x}_r-\boldsymbol{x}_\ell)\cdot(\boldsymbol{\nabla}\boldsymbol{u})_r
\end{equation}
\begin{equation}
p_L = p_\ell + \alpha\,(\boldsymbol{x}_r-\boldsymbol{x}_\ell)\cdot(\boldsymbol{\nabla}p)_\ell\;,
\quad
p_R = p_r - \alpha\,(\boldsymbol{x}_r-\boldsymbol{x}_\ell)\cdot(\boldsymbol{\nabla}p)_r
\end{equation}

\noindent
Here, the coefficient $\alpha$ is set to $0.36$ to optimise the dispersive properties of the scheme, and the gradients are evaluated using the classic Green-Gauss approximation. According to Löwe et al. \cite{lowe2016low}, the extrapolation of the density and the speed of sound has negligible impact on subsonic flows and is therefore omitted in the current implementation (i.e., $c_L=c_\ell$, $c_R=c_r$, $\rho_L=\rho_\ell$, $\rho_R=\rho_r$).

Since the LD2 scheme is designed to produce very low numerical dissipation, its basic formulation should not be applied throughout the computational domain \cite{lowe2016low}. Indeed, the lack of dissipation is likely to cause numerical instabilities in regions with poor mesh quality (e.g., with highly skewed cells). To avoid this, Probst and Reuß \cite{probst2015scale} proposed a hybrid formulation that blends the LD2 scheme with a more robust and dissipative reference discretisation:

\begin{equation}
\boldsymbol{\Tilde{F}}_{LR} = (1-\sigma)\boldsymbol{F}_{LR} + \sigma \boldsymbol{F}_{LR}^{ref}
\label{eq:blending_old}
\end{equation}

\noindent
In the above expression, $\boldsymbol{F}_{LR}^{ref}$ denotes the flux computed using the aforementioned reference scheme, and $\sigma$ is the blending function introduced by Travin and Shur \cite{travin2002physical} to distinguish between rotational and irrotational regions of the flow field.
In the present work, the same approach as in Eq. \eqref{eq:blending_old} is employed but the modified shielding function $\tilde{f}_d$ is used to perform the blend in place of $\sigma$. Futhermore, for the sake of consistency with the discretisation of the convective terms in the turbulence model equations, a first-order upwind scheme has been selected to compute $\boldsymbol{F}_{LR}^{ref}$ ($\equiv\boldsymbol{F}_{LR}^{upw}$). In summary, the convective fluxes at cell-face centres can be expressed as:

\begin{equation}
\boldsymbol{\Tilde{F}}_{LR} = \Tilde{f}_d\,\boldsymbol{F}_{LR} + (1-\Tilde{f}_d) \boldsymbol{F}_{LR}^{upw}
\label{eq:blending}
\end{equation}

\vfill
\noindent
This choice was motivated by two main considerations: first, the LD2 scheme, which requires the evaluation of the pressure gradients (not used elsewhere in the code), is applied only when really necessary (i.e., when the LES mode is activated), thus reducing the overall computational cost of the methodology. In addition, Eq. \eqref{eq:blending} prevents potential inconsistencies between the definitions of $\sigma$ and $\Tilde{f}_d$. In fact, $\sigma$ is designed to identify vortical structures throughout the computational domain, tending to zero both in near-wall RANS zones (where $\tilde{f}_d\approx0$) and in outer irrotational flow regions (where, instead, $\tilde{f}_d\approx1$).

It should be emphasised that the blend described in Eq. \eqref{eq:blending} is not generally applied everywhere. The multi-block structure of UZEN flow solver enables a zonal specification of the simulation methodology. Specifically, the user can select the blocks where X-LES ($+$ SBS) is to be activated, while the remaining blocks are automatically treated in RANS mode.
In RANS blocks, the convective terms are discretised using the original second-order central differencing scheme, supplemented with JST-type scalar artificial dissipation. Obviously, in X-LES blocks, the blend of the LD2 scheme with an upwind discretisation eliminates the need for any additional artificial diffusion.
Nonetheless, the user is also given the possibility to enforce the pure LD2 scheme (without any blend), in combination with a small amount of numerical dissipation. For further details on this aspect, the reader is referred to \cite{carlsson2023investigation}. 

\begin{figure}[h]
\centering
\sidesubfloat[]{\includegraphics[width=0.45\linewidth,frame]{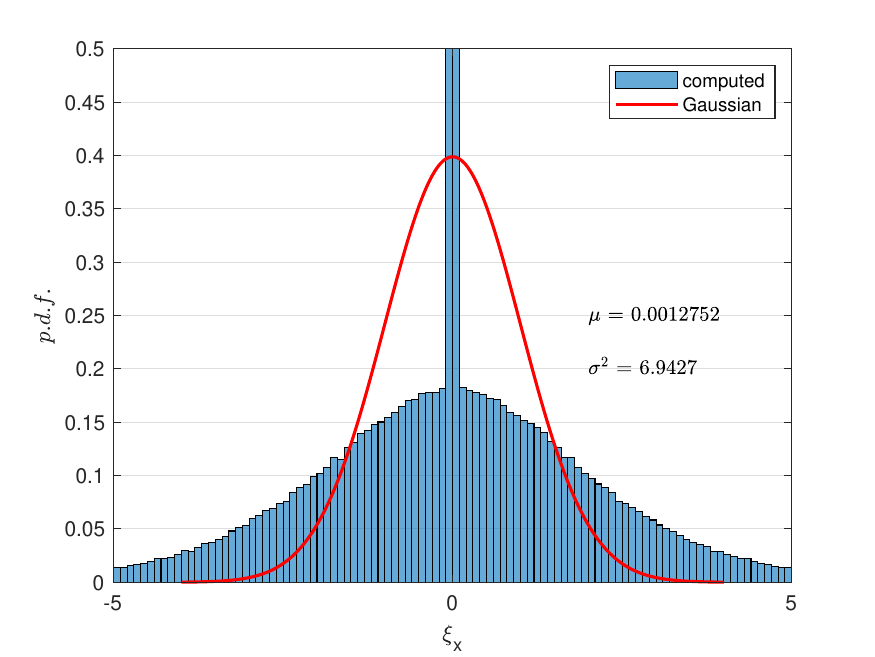}\label{fig:csi:a}}
\hfill
\sidesubfloat[]{\includegraphics[width=0.45\linewidth,frame]{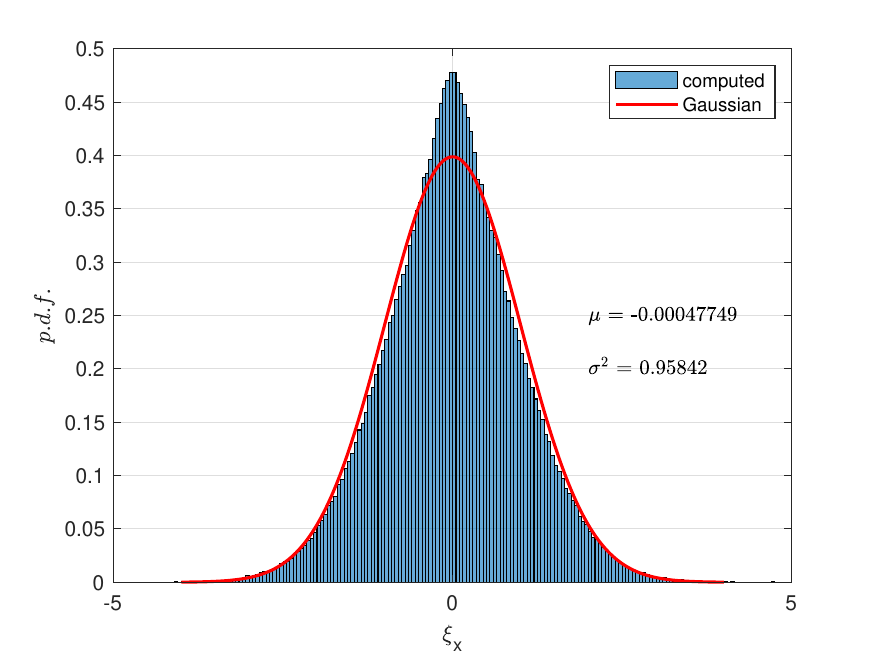}\label{fig:csi:b}}
\caption{Probability density function of the $x$-component of the stochastic variable $\boldsymbol\xi$ in an X-LES block from the wake-boundary-layer mixing test case:  first-order upwind scheme (a) vs. blended LD2–upwind scheme (b).}
\end{figure}

The use of a central scheme for the discretisation of the convective terms in the balance equations is also critical for the preservation of the stochastic properties in the Langevin-type equations. As shown by Kok \cite{kok2017stochastic}, non-central schemes fail to preserve the variance of the variable $\boldsymbol{\xi}$. This is clearly shown in Figs. \ref{fig:csi:a}–\ref{fig:csi:b}, which report the probability density function of the $x$-component of the latter vector quantity for a representative X-LES block in the wake-boundary-layer mixing test case (see Section \ref{subsec:mixing}).
As expected, the upwind scheme (Fig. \ref{fig:csi:a}) both increases the variance of $\xi_x$ and introduces numerical damping, resulting in a peak around $\xi_x = 0$. In contrast, the blended LD2-upwind scheme (Fig. \ref{fig:csi:b}) preserves the Gaussian shape of the distribution almost perfectly.

\subsection{Wall distance}
\label{subsec:wall_dist}

The shielding function used for Delayed X-LES depends on the distance to the nearest wall (see Eq. \eqref{eq:shielding}). 
The algorithm for the evaluation of the latter quantity across the computational domain must be handled carefully, since it must ensure continuity and sufficient robustness in the presence of multiple closely spaced walls.
A method that meets these criteria is based on solving a Poisson equation for the so-called wall distance variable, $\phi$ \cite{spalding1994calculation,tucker2005computations}:

\begin{equation}
    \nabla^2\phi = -1
    \label{eq:Poisson}
\end{equation}

\noindent
Homogeneous Dirichlet boundary conditions are enforced at solid walls, whereas homogeneous Neumann conditions are imposed at the remaining boundaries.
The wall distance variable is related to the true wall distance through the following normalisation formula:

\begin{equation}
    d_w = \sqrt{\boldsymbol{\nabla}\phi\,\cdot\,\boldsymbol{\nabla}\phi\,+2\phi} - |\boldsymbol{\nabla}\phi|
\end{equation}
A rigorous mathematical justification for the above equation is provided by Fares \& Schr\"oder \cite{fares2002differential}. In the present dissertation, only the numerical aspects of the methodology are discussed. Eq. \eqref{eq:Poisson} is solved in a finite-difference framework, using cell centres as computational nodes.
Its discrete counterpart can be expressed as:

\begin{equation}
    L\,\varphi = b
    \label{eq:sys}
\end{equation}

\noindent
where $L$ denotes the discrete Laplace operator, $\varphi$ the discrete form of the continuous variable $\phi$ and $b$ a vector of ones. The linear system in Eq. \eqref{eq:sys} is solved using the Accelerated Over-Relaxation (AOR) \cite{hadjidimos1978accelerated} method. This requires the Laplace operator to be decomposed as:

\begin{equation}
    L = L_D+L_U+L_L
\end{equation}

\noindent
with $L_U$ and $L_L$ being the upper and lower triangular parts of $L$, respectively, and $L_D =\mathrm{diag}(L)$.
Letting $\varphi^{(n)}$ denote the value of the discrete variable $\varphi$ in the n-th iteration, the AOR algorithm assumes the following form:

\begin{equation}
    \left[I-r\left(L_D^{-1}L_L\right)\right] \varphi^{(n)} = \left[(1-\omega_d)I+(\omega_d-r)(L_D^{-1}L_L)+\omega_d(L_D^{-1}L_U)\right]\,\varphi^{(n-1)} + \omega_d (L_D^{-1}\,b)
    \quad
    | \quad
    n=1,...
\end{equation}

\noindent
where $r$ and $\omega_d$ are two fixed scalar parameters. In particular, when $r=\omega_d$, the algorithm reduces to the well-known Successive Over-Relaxation (SOR) method.

\begin{figure}[h]
    \centering
    \sidesubfloat[]{\includegraphics[width=0.45\linewidth,frame]{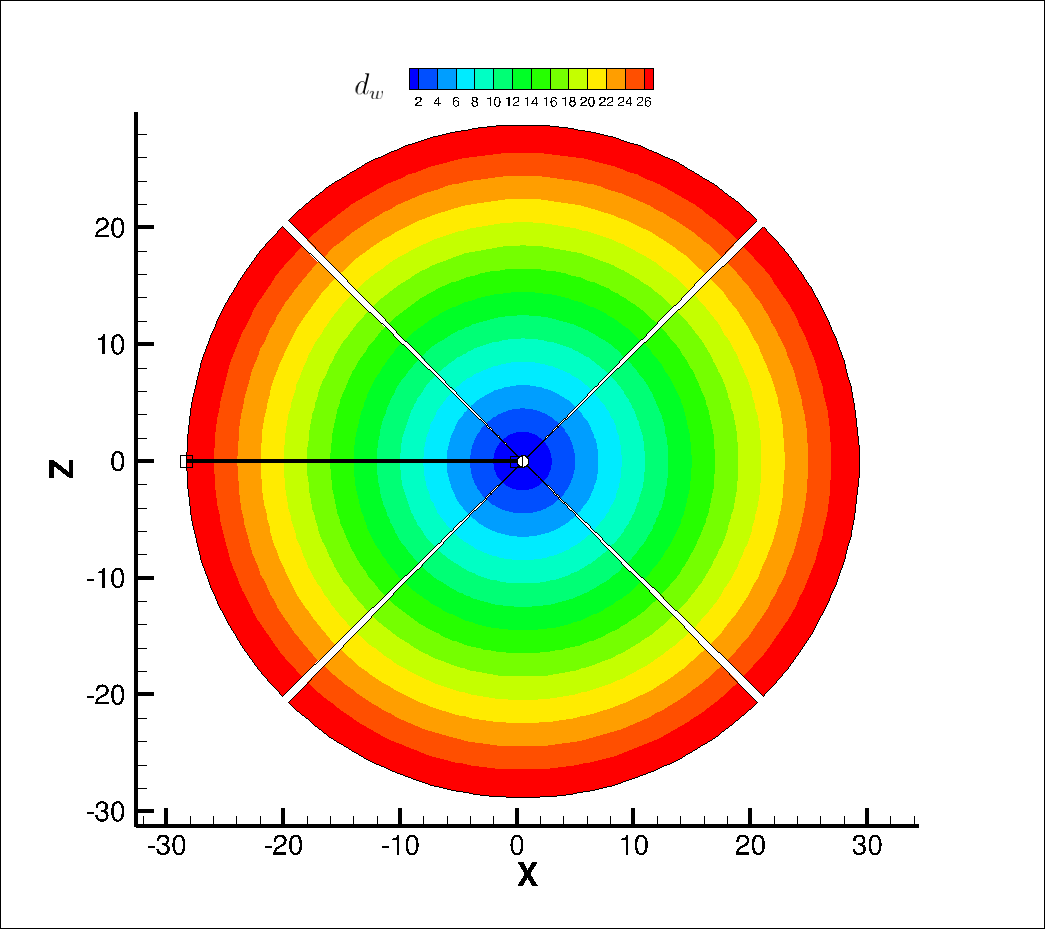}\label{fig:dist_cyl:a}}
    \hfill
    \sidesubfloat[]{\includegraphics[width=0.45\linewidth,frame]{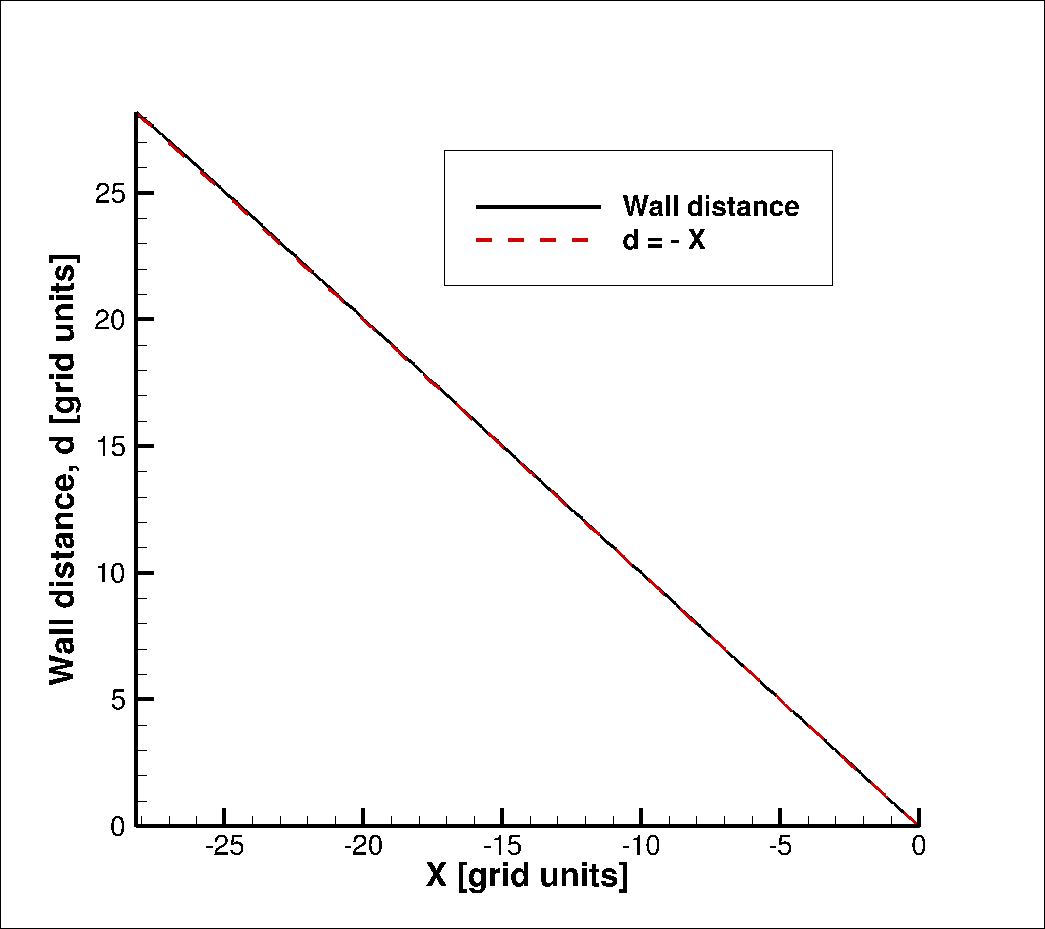}\label{fig:dist_cyl:b}}
    \caption{Wall distance field for the cylinder test case: contour (a) and plot along a reference line (b). The reference line (solid black, square markers at its ends) is shown in (a).}
\end{figure}

\begin{figure}[h]
    \centering
    \sidesubfloat[]{\includegraphics[width=0.45\linewidth,frame]{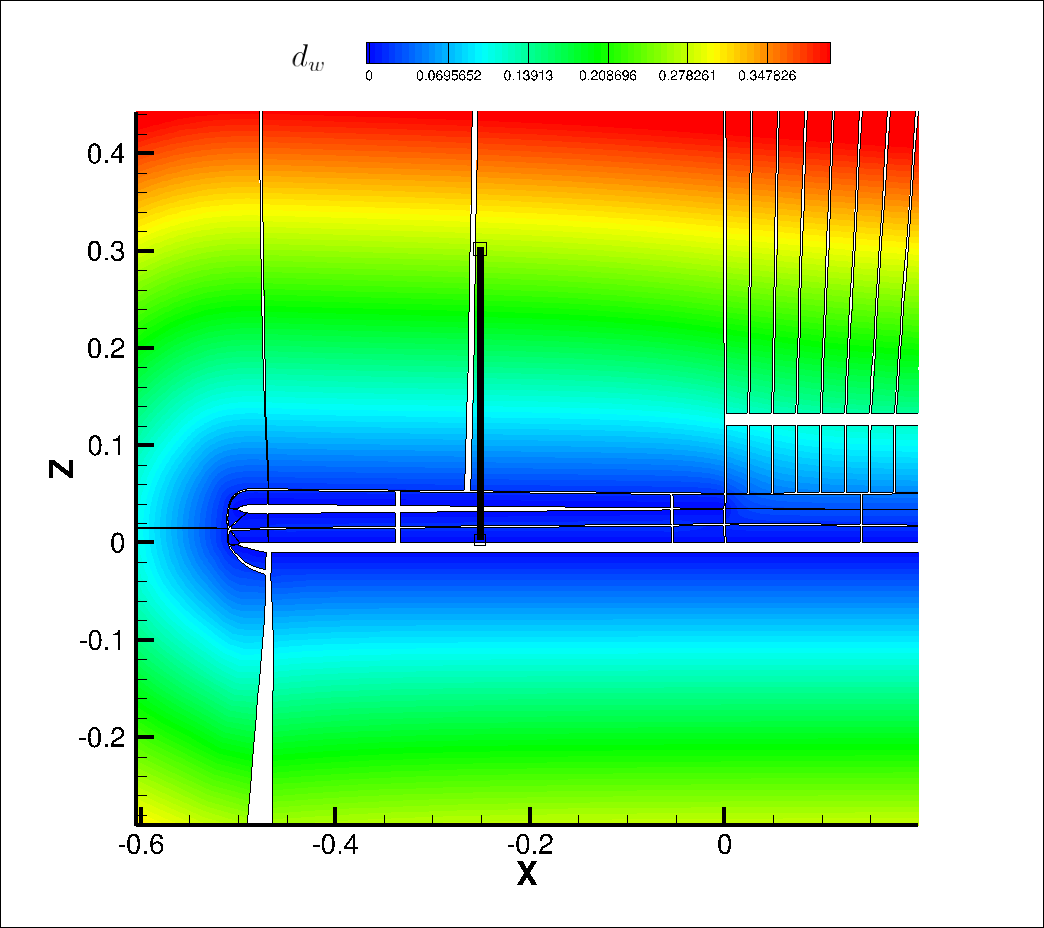}\label{fig:dist_mixing:a}}
    \hfill
    \sidesubfloat[]{\includegraphics[width=0.45\linewidth,frame]{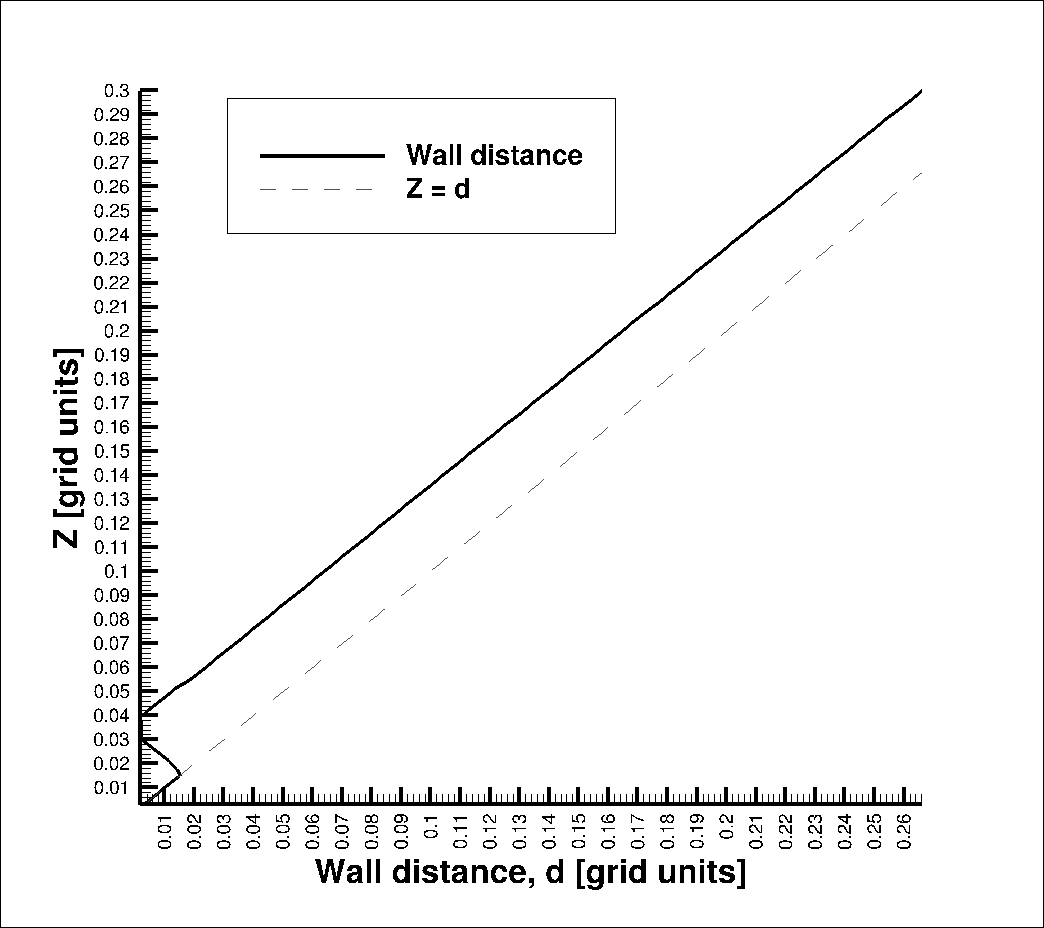}\label{fig:dist_mixing:b}}
    \caption{Wall distance field for the wake-boundary-layer mixing test case (see Section \ref{subsec:mixing}): contour (a) and plot along a reference line (b). The reference line (solid black, square markers at its ends) is shown in (a).}
\end{figure}

The accuracy and reliability of the algorithm are verified for two geometric test cases of increasing complexity: a circular right cylinder enclosed in a cylindrical computational domain (Figs. \ref{fig:dist_cyl:a}–\ref{fig:dist_cyl:b}), and a symmetric airfoil positioned above a flat plate (Figs. \ref{fig:dist_mixing:a}–\ref{fig:dist_mixing:b}). In both cases, the resulting distance fields are found to be satisfactory. In particular, in the second configuration, the wall distance along the solid black line reported in Fig. \ref{fig:dist_mixing:a}, extending from the upper surface of the flat plate to an unspecified point above the airfoil, exhibits a distinct corner in the segment between the two bodies. As expected, this corner appears exactly halfway between the plate and the airfoil.
Fig. \ref{fig:dist_mixing:b} also includes the bisector line to confirm that the slope of the $d_w$-curve is the correct one.
\vfill

\vfill 

\subsection{Stochastic equations}
\label{subsec:stoc_eqs}
The discretisation of the stochastic equations introduced in Section \ref{subsec:SBS} is discussed below.
As anticipated, the source term in the Langevin equations is a spatially correlated stochastic differential (denoted as $dW_i$ in Eq. \eqref{eq:Langevin}).
This can be computed through the implicit smoothing of a completely uncorrelated stochastic differential.
Therefore, the latter is the true starting point for the integration of the Langevin equations. The solution procedure consists of three consecutive steps:
\vspace{10pt}

\begin{enumerate}
    \item \textit{Generation of uncorrelated stochastic variables}\\[2pt]
    A vector of three independent, normally-distributed stochastic variables is generated at each grid cell. The components of this vector must be uncorrelated both in space and in time. Therefore, they are independently sampled at each time step. They represent the discrete equivalent of the stochastic differential $dV_i$ in Eq.~\eqref{eq:dVi} and are denoted as $(\zeta_m)_{ijk}^n$, where $n$ refers to the $n$-th time instance, and $ijk$ to the cell indices in the computational space. Each component $\zeta_m$ must follow a Gaussian distribution with zero mean and unit variance:

    \begin{equation}
        (\zeta_m)_{ijk}^n {=} \mathcal{N}(0,1)
        \quad
        m=1,2,3
    \end{equation}

    This requirement is met by using the Box–Muller transform~\cite{box1958note}, which maps two uniformly-distributed random variables, belonging to the interval $(0,1)$, to two standard normally-distributed samples. Since the transform produces two stochastic variables per each application, it needs to be applied only $N_{\scalebox{0.5}{XLES}}/2$ times per time step, where $N_{\scalebox{0.5}{XLES}}$ is the total number of cells in the X-LES blocks.

    \vspace{10pt}
    \item \textit{Implicit smoothing of the uncorrelated stochastic variables}\\[2pt]
    An implicit smoothing is applied to the uncorrelated stochastic variables $(\zeta_m)_{ijk}^n$ to ensure the desired exponentially-decaying spatial correlation (see Eq.~\eqref{eq:smooth}):

    \begin{equation}
          \left(I - C_\Delta \frac{\Delta^2}{\delta x_k^2} \delta_k^2\right)\left(I - C_\Delta \frac{\Delta^2}{\delta x_j^2} \delta_j^2\right)\left(I - C_\Delta \frac{\Delta^2}{\delta x_i^2} \delta_i^2\right) \frac{(\eta_m)_{ijk}^n}{\lambda_{ijk}} = (\zeta_m)_{ijk}^n
        \quad
        m=1,2,3
        \label{eq:disc_smooth}
    \end{equation}

    In the above equation, $\delta x_p$ (with $p=i,j,k$) is the mesh spacing in the $p$-th direction, whereas $\delta_p^2$ denotes the second-order central differencing operator. The correction factor $\lambda_{ijk}$ is introduced to preserve the variance of the smoothed field, i.e., to ensure that $\mathrm{Var}\{(\eta_m)_{ijk}^n\}=\mathrm{Var}\{(\zeta_m)_{ijk}^n\}=1$ \cite{kok2017stochastic}. It is defined as:

    \begin{equation}
        \lambda_{ijk} = \frac{(1 + 4\beta_i)^{3/4}(1 + 4\beta_j)^{3/4}(1 + 4\beta_k)^{3/4}}{(1 + 2\beta_i)^{1/2}(1 + 2\beta_j)^{1/2}(1 + 2\beta_k)^{1/2}}
    \end{equation}
    \begin{equation*}
        \beta_i = C_\Delta \frac{\Delta^2}{\delta x_i^2},
        \quad
        \beta_j = C_\Delta \frac{\Delta^2}{\delta x_j^2},
        \quad
        \beta_k = C_\Delta \frac{\Delta^2}{\delta x_k^2}
    \end{equation*}

    Since the system in Eq. \eqref{eq:disc_smooth} is tridiagonal, the implicit smoothing can be carried out through a triple application of the Thomas algorithm. Homogeneous Dirichlet boundary conditions are enforced at the block external boundaries.

    \vspace{10pt}
    \item \textit{Integration of the Langevin equations} \\[2pt]
    The spatially and temporally correlated stochastic variables $\xi_m$ are obtained by integrating Eq.~\eqref{eq:Langevin} in time, using a second-order backward Euler scheme (which is the default temporal discretisation in UZEN flow solver):

    \begin{equation}
        (\rho \xi_m)_{ijk}^{n} + \frac{\tau}{2\,\delta t} \left[3(\rho \xi_m)_{ijk}^{n} - 4(\rho \xi_m)_{ijk}^{n-1} + (\rho \xi_m)_{ijk}^{n-2}\right] + \tau\, C_{ijk}^n = F_c \sqrt{\frac{2 \tau}{\delta t}}\, \rho_{ijk}^n\, (\eta_m)_{ijk}^n
        \label{eq:Langevin_disc}
        \quad
        m=1,2,3
    \end{equation}

    \vspace{2pt}
    Here, $\delta t$ is the physical time step, and $C_{ijk}$ the convective term discretised using the LD2 scheme (see Section~\ref{subsec:convec}).  
    Since variance preservation is only ensured when central schemes are employed~\cite{kok2017stochastic} (both in space and in time), a correction factor $F_c$ is introduced on the right-hand side of Eq.~\eqref{eq:Langevin_disc}:

    \begin{equation}
        F_c = \sqrt{\frac{(1+a)(4+a)}{2(2+a)}},
        \quad
        a \equiv \frac{\delta t}{\tau}
    \end{equation}

    This was obtained using an analysis similar to the one reported in~\cite{kok2017stochastic} (see the Appendix for further details).
    To conclude, it must be stressed that the right-hand side of Eq.~\eqref{eq:Langevin_disc} is set to zero in RANS mode. As a result, the stochastic variables decay exponentially in a Lagrangian sense when $\tilde{f}_d = 0$.
\end{enumerate}

\section{Results}
\label{sec:results}

\subsection{Decaying Isotropic Homogeneous Turbulence (DIHT)}
\label{subsec:calib}
From the calibration of the SBS model on the classic decay of isotropic homogeneous turbulence, Kok \cite{kok2017stochastic} obtained a model constant of $C_1 = 0.08$ using a low-dissipation low-dispersion fourth-order scheme \cite{kok2009high}. Since $C_1$ is known to be sensitive to the numerical discretisation employed, the calibration is repeated in CIRA's flow solver.

Following Kok \cite{kok2017stochastic}, the computations are performed on a $64^3$ grid, generated within a cubic domain of size $L = 11M$, where $M = \SI{5.08}{\centi\meter}$ corresponds to the spacing in the (metal) grid used by Comte-Bellot \& Corrsin for their well-known experiment \cite{comte1966use}. In their setup, the inflow velocity was $U_0 = \SI{10}{\meter\per\second}$, yielding a Reynolds number of $Re_0 = U_0 M / \nu = \SI{34000}{}$.
Periodic boundary conditions are applied in each direction of the computational domain.
The initial velocity field is generated to match the experimental energy spectrum at time $t^+ = t U_0 / M = 42$. This is accomplished using TurboGenPY, a synthetic isotropic turbulence generator specifically designed for constant-density flows. The tool ensures a discrete divergence-free condition on structured grids and reproduces the target spectrum up to the Nyquist limit \cite{saad2017scalable,saad2016comment,richards2018fast}.
The initial density field is uniform and consistent with the free-stream conditions in the experiment. Since the flow is nearly incompressible, the initial pressure field can be computed after the generation of the random velocity field by simply solving a Poisson equation:
\begin{equation}
\begin{split}
    \dfrac{\partial p}{\partial x_i \partial x_i}&=-\rho\,\dfrac{\partial u_i}{\partial x_j}\dfrac{\partial u_j}{\partial x_i}
\end{split}
\end{equation}
Simulations are performed in full LES mode (i.e., $\Tilde{f}_d = 1$ everywhere). Regarding the turbulence variables, an initial condition for the subgrid kinetic energy is needed. This is prescribed by assuming a balance between the production and the dissipation of $k$ \cite{kok2004extra}:

\begin{equation}
k = \dfrac{1}{\beta_k} (C_1 \Delta)^2\,S_{ij}\,S_{ij}
\end{equation}

\noindent
where $S_{ij}=\partial u_i/\partial x_j$ is the rate-of-strain tensor and $\beta_k=0.09$.
For the present test case, high-pass filtering in time is disabled. In addition, due to the near-incompressibility of the flow, a preconditioning method is employed \cite{puoti2003preconditioning}.

\begin{figure}[h]
\centering
\sidesubfloat[]{\includegraphics[width=0.45\linewidth,frame]{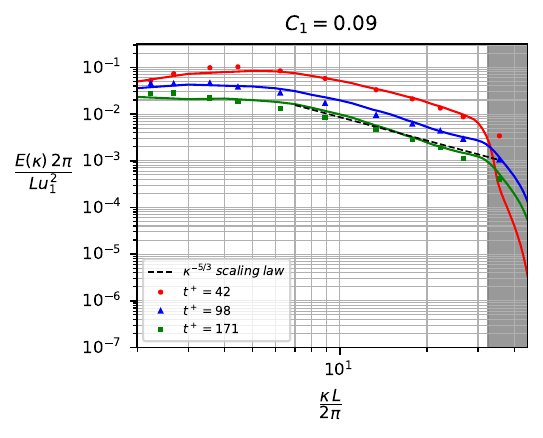}\label{fig:calib:a}}
\hfill
\sidesubfloat[]{\includegraphics[width=0.45\linewidth,frame]{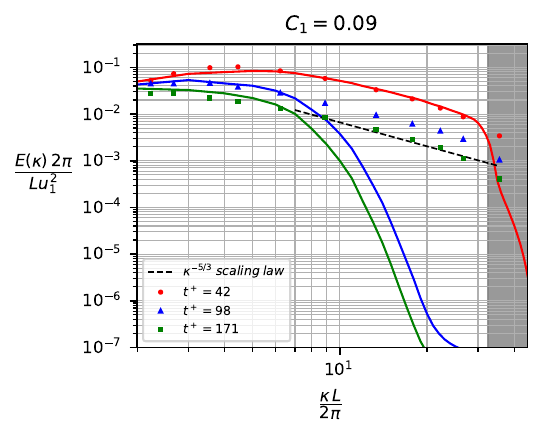}\label{fig:calib:b}}
\caption{Turbulent energy spectra at three time instances ($\boldsymbol{t^+ = 42, 98, 171}$): LD2 scheme (a) vs. JST scheme (b). Experimental data are represented by markers; solid lines denote computational results.}
\end{figure}

Through a least-squares optimisation process, a model constant of $C_1 = 0.09$ is obtained. The latter is only slightly greater than the one reported by Kok \cite{kok2017stochastic}. The resulting non-dimensional spectra are shown in Fig. \ref{fig:calib:a}. $u_1$ denotes the r.m.s. fluctuating velocity computed from the first experimental spectrum ($t^+ = 42$):

\begin{equation}
u_1 = \sqrt{\dfrac{2}{3}\left(\int_0^\infty E(\kappa)\big\rvert_{t^+ = 42}\;d\kappa\right)} = \SI{0.222}{\meter\per\second}
\end{equation}

\noindent
The $-5/3$ Kolmogorov scaling law is also shown for the latest time instance.
The shaded area in the plot corresponds to wavenumbers beyond the non-dimensional Nyquist limit, $\kappa^+_{\max}$:

\begin{equation}
\kappa_{\max}^+ = \kappa_{\max}\cdot\dfrac{L}{2\pi} = \dfrac{\pi}{\Delta} \cdot \dfrac{L}{2\pi} = \dfrac{L}{2(L/64)} = 32
\end{equation}

\noindent
The value of $C_1$ obtained from this calibration will be used in the remainder of this work.

For the sake of completeness, the turbulent spectra computed using the JST scheme (with the same model constant, $C_1=0.09$) are shown in Fig. \ref{fig:calib:b}. It is evident that the highly dissipative nature of this type of discretisation prevents an adequate resolution of the smallest turbulent scales.

\subsection{Wake-boundary-layer mixing}
\label{subsec:mixing}

\begin{figure}
    \centering
    \sidesubfloat[]{\includegraphics[width=0.45\linewidth,frame]{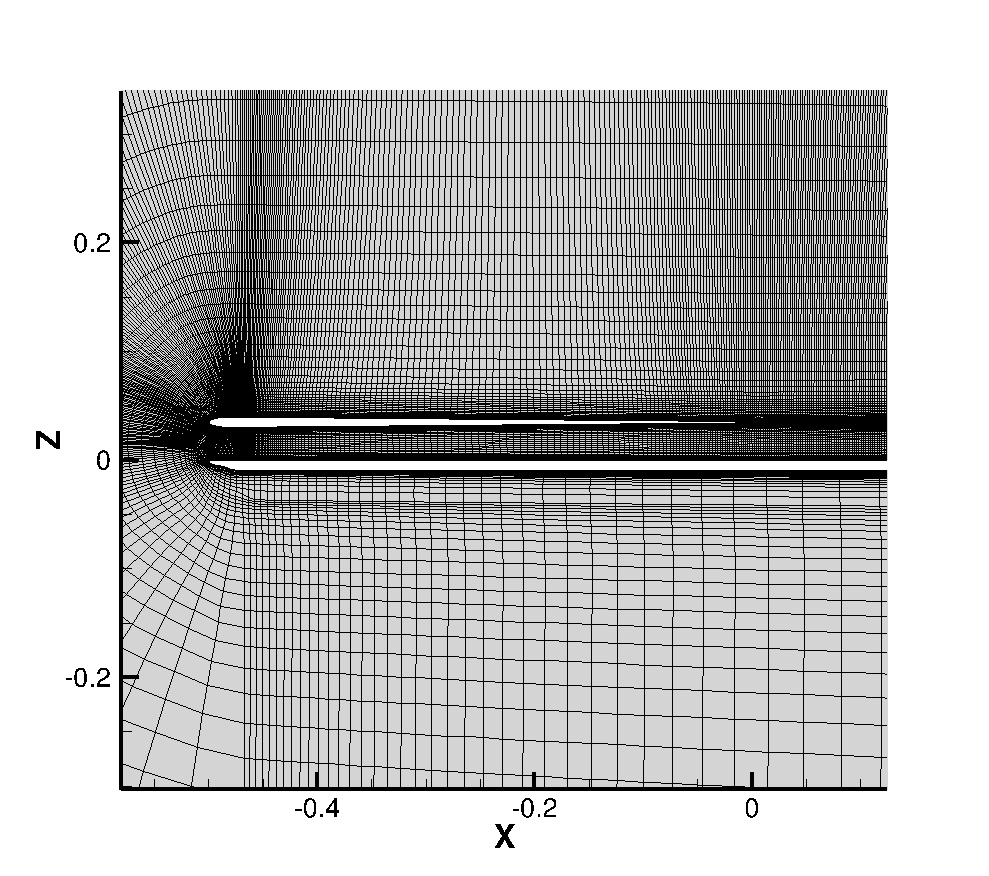}\label{fig:grid:a}}
    \hfill
    \sidesubfloat[]{\includegraphics[width=0.45\linewidth,frame]{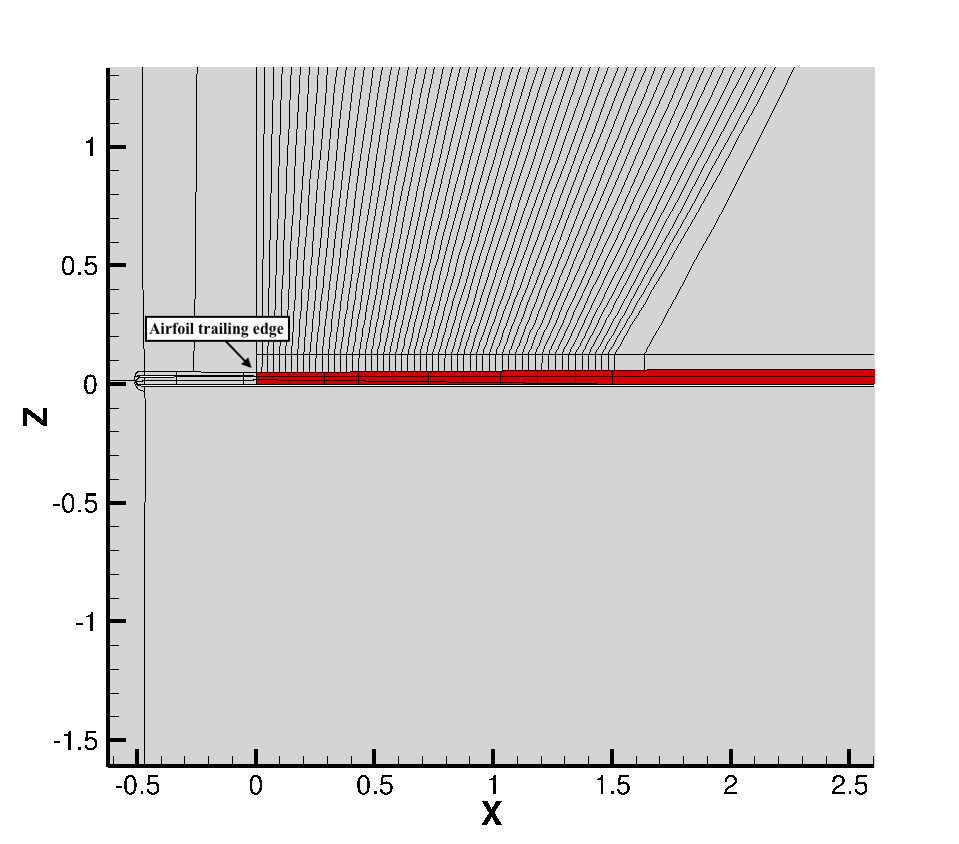}\label{fig:grid:b}}
    \caption{Grid for the wake-boundary-layer mixing test case (a). Blocks simulated by X-LES + SBS in (b).}
\end{figure}

To demonstrate the effectiveness of the SBS model in mitigating the grey area issue, the wake-boundary-layer mixing test case is considered. The latter consists in the mixing between the wake past an airfoil at zero incidence and the zero-pressure-gradient turbulent boundary layer developing over a flat plate.
Such a configuration serves as a simplified representation of the complex aerodynamic interference effects which typically arise in multi-element airfoil systems.
Experimental investigations on this setup were carried out at the Netherlands Aerospace Centre (NLR) in 1979 \cite{pot1979wake} and their outcome is used here for validation purposes.
The computational grid employed is partially shown in Fig. \ref{fig:grid:a}. It is made of $171$ blocks, for a total of about $10^7$ control volumes. The reference length ($L_{\scalebox{0.6}{ref}}$) is equal to the airfoil chord ($c=\SI{500}{\milli\meter}$). 
The domain size in the spanwise direction ($Y$ axis in Fig. \ref{fig:grid:a}) is $15\%\,L_{\scalebox{0.6}{ref}}$ long and is equally divided into $72$ cells.
The distance between the airfoil trailing edge and the flat plate is $H=7\%\,L_{\scalebox{0.6}{ref}}=\SI{35}{\milli\meter}$.
The freestream Reynolds number per unit length is $Re_{\infty}/L_{\scalebox{0.6}{ref}} = 2.4\cdot10^6\,\SI{}{\per\meter}$, whereas the Mach number is $M_\infty=0.08$. 
The 33 blocks which are simulated in X-LES mode (+ SBS) are highlighted in red in Fig. \ref{fig:grid:b}; the remaining blocks are permanently treated in RANS mode. 
The filter width is fixed; in particular, it is set equal to the (uniform) cell width in the $Y$ direction (i.e., $0.15\cdot L_{\scalebox{0.6}{ref}}/72 = \SI{1.042}{\milli\metre}$). The simulation is advanced in time with a physical time step of about $3.6\cdot10^{-7}\,\SI{}{\second}$.
The CFL number selected for inner-time integration is 0.9. This represents the maximum value which ensures the stability of the numerical method for the present test case. No residual averaging nor multistage methods are applied.
Convergence in dual time is achieved after about $40$ iterations. The convergence criterion consists in the reduction of the r.m.s. density residual of two orders of magnitude.

\begin{figure}[h]
    \centering
    \sidesubfloat[]{\includegraphics[width=0.45\linewidth,frame]{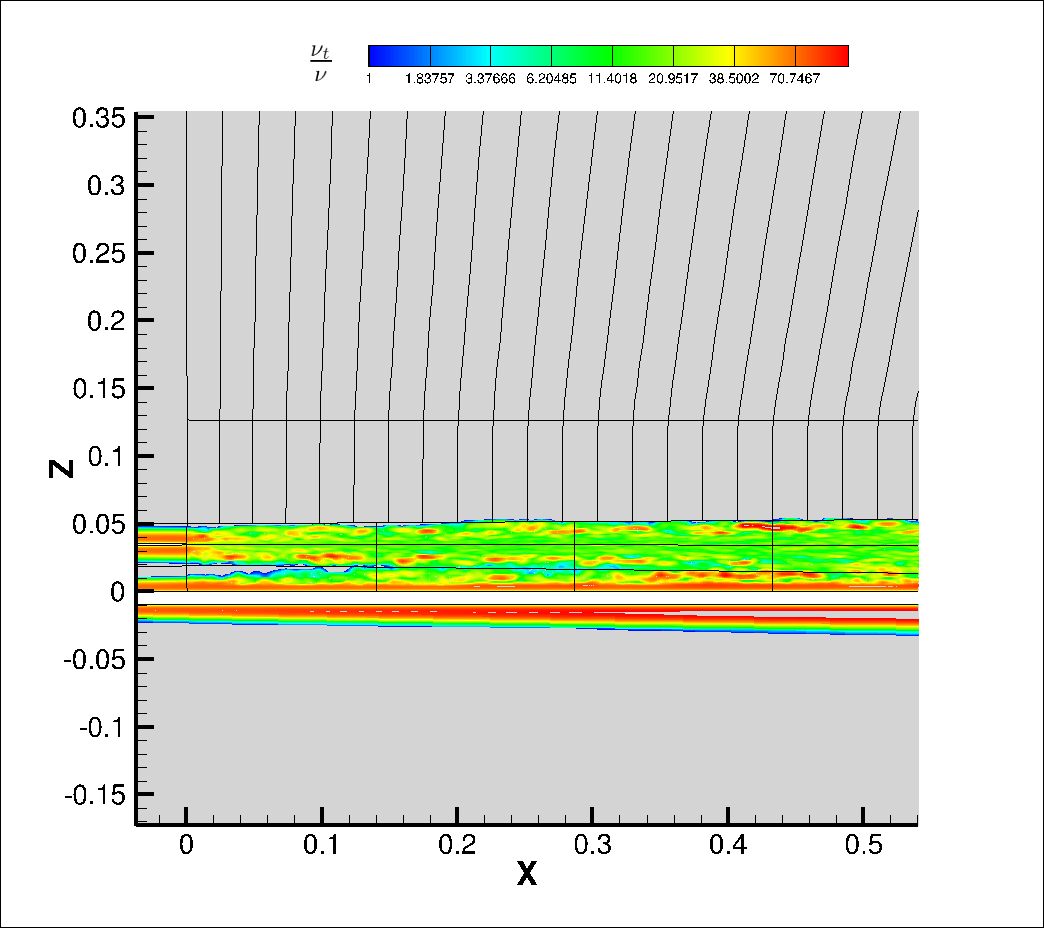}\label{fig:eddy:a}}
    \hfill
    \sidesubfloat[]{\includegraphics[width=0.45\textwidth,frame]{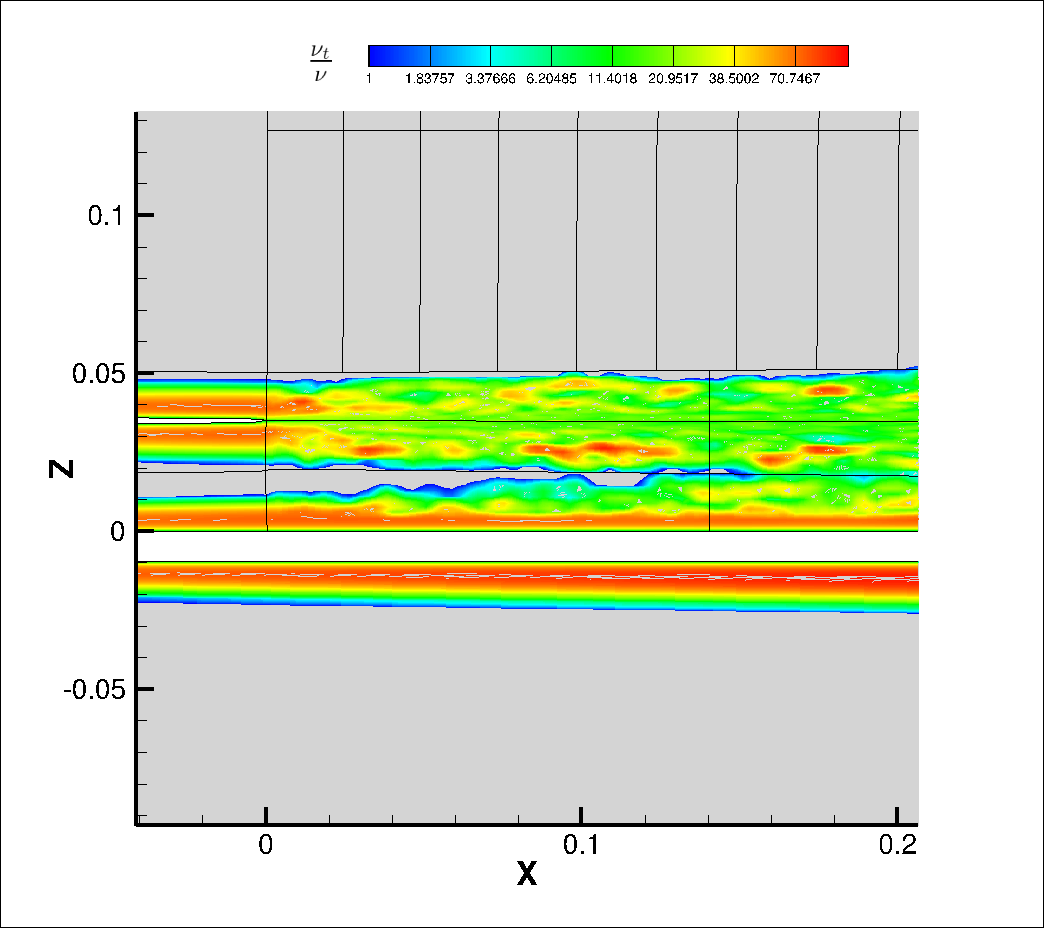}\label{fig:eddy:b}}
    \caption{Modelled viscosity ratio ($\nu_t/\nu$) in X-LES blocks (a) and close-up on the airfoil trailing edge (b).}
\end{figure}

\begin{figure}[h]
    \centering
    \sidesubfloat[]{\includegraphics[width=0.45\linewidth,frame]{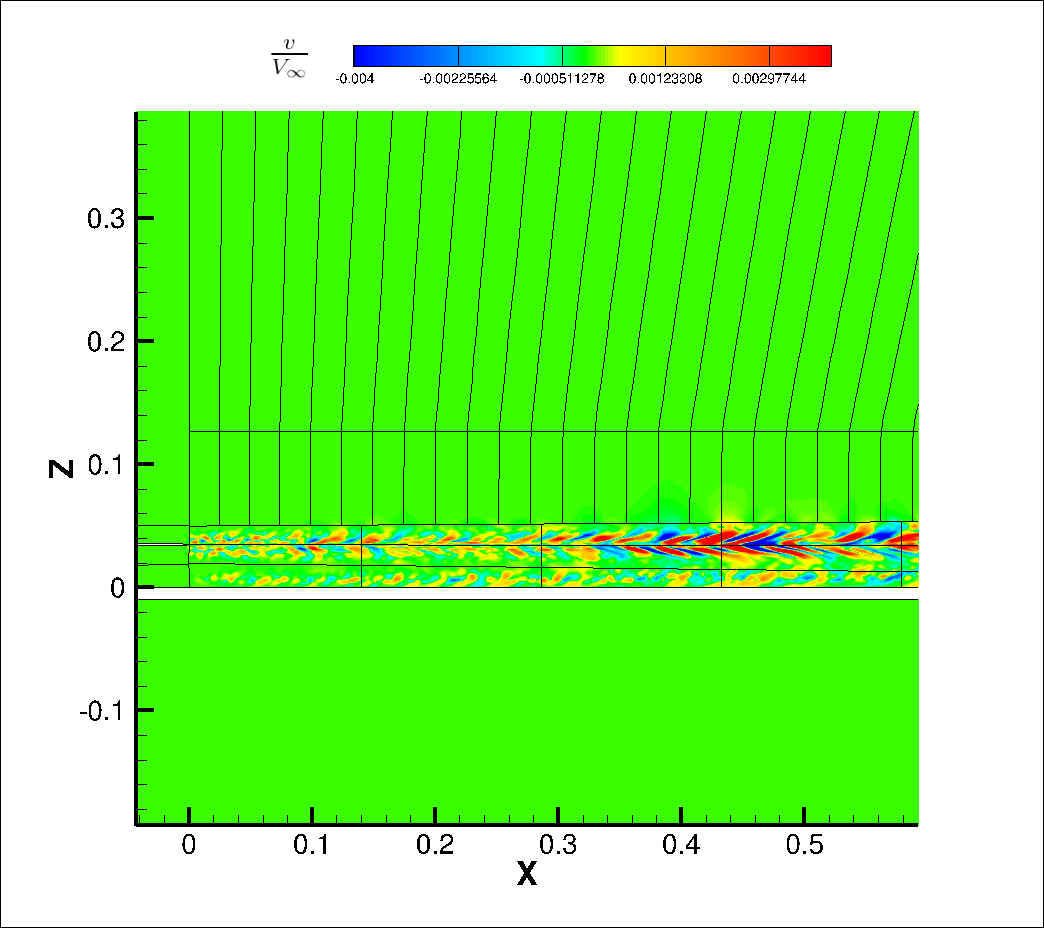}\label{fig:vy:a}}
    \hfill
    \sidesubfloat[]{\includegraphics[width=0.45\textwidth,frame]{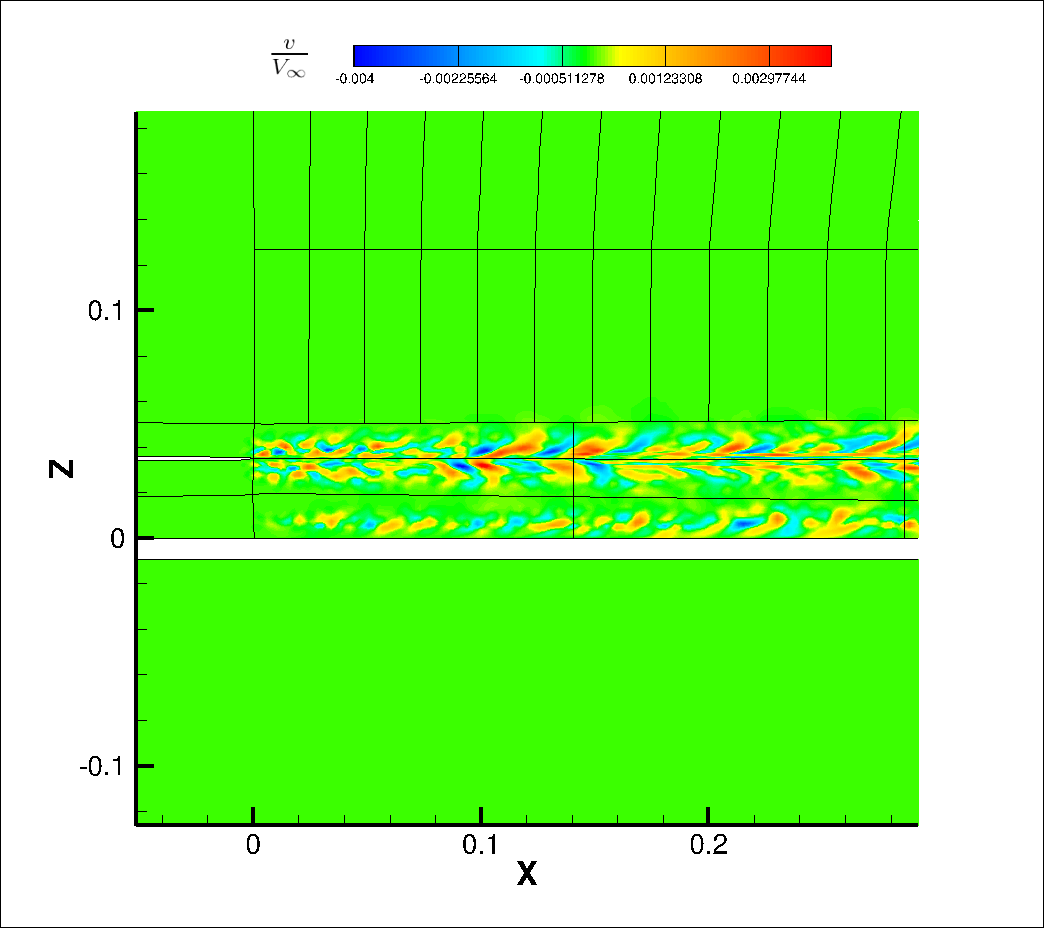}\label{fig:vy:b}}
    \caption{Non-dimensional spanwise velocity in X-LES blocks (a) and close-up on the airfoil trailing edge (b).}
\end{figure}

\begin{figure}[h]
    \centering
    \includegraphics[width=0.6\linewidth,frame]{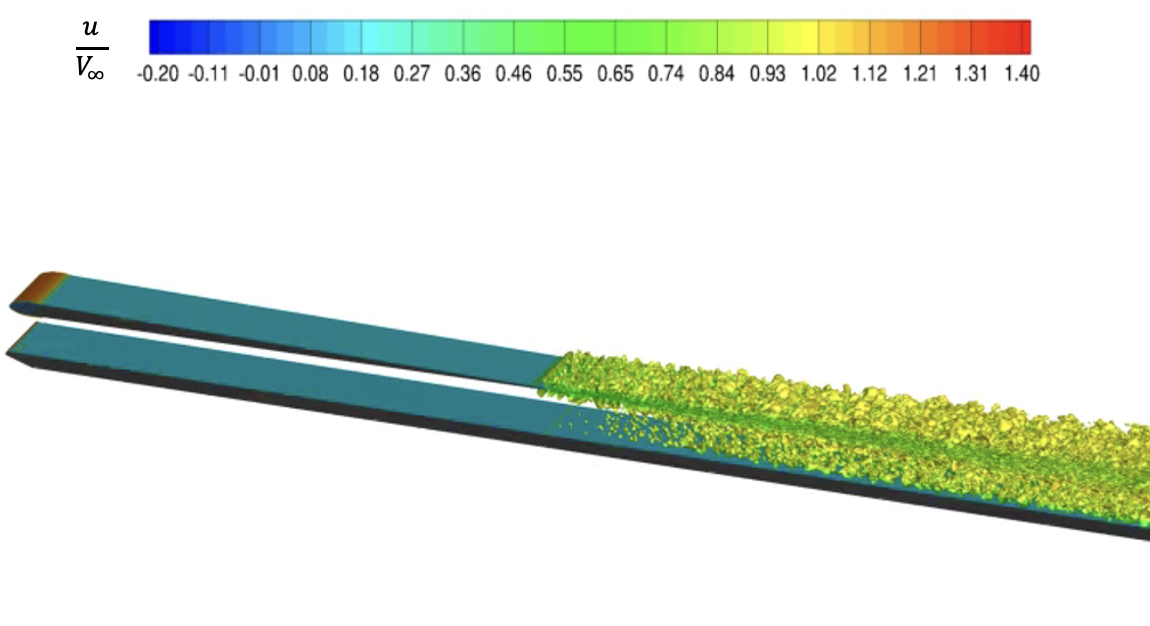}
    \caption{Iso-surface of Q-criterion ($QL_{\scalebox{0.6}{ref}}^2/V_\infty^2=50$) from the instantaneous flow field colored by the non-dimensional $X$-velocity component.}
    \label{fig:q_crit}
\end{figure}

As shown in Figs. \ref{fig:eddy:a}-\ref{fig:eddy:b}, the flow experiences a sudden reduction in the eddy/subgrid viscosity when transitioning from the RANS blocks, which are adjacent to the airfoil upper and lower surfaces, to the X-LES blocks, that are located within the airfoil wake.
This is associated with a decrease in the modelled part of the Reynolds stress tensor, an occurrence which is expected to trigger the formation of fluctuating turbulent structures. 
In fact, looking at the contours of the spanwise velocity component shown in Figs. \ref{fig:vy:a}-\ref{fig:vy:b}, it can be noticed that the destabilisation of the free shear layer takes place exactly at the airfoil trailing edge (namely, at the front boundaries of the first X-LES blocks along the $X$ axis).
This can also be observed more clearly in Fig. \ref{fig:q_crit}, where an iso-surface of Q-criterion from the instantaneous flow field is shown.
It can be concluded that the grey area has efficiently been removed.
The improvement with respect to previous version of the model, i.e., the Stochastic Eddy Viscosity (SEV) \cite{kok2010destabilizing}, is evident. 
Indeed, Catalano \cite{catalano2019application} applied the SEV method to the same test case and observed that fluctuating turbulent structures started to develop at a distance of almost \SI{250}{\milli\meter} ($=c/2$) from the airfoil trailing edge. 

\begin{figure}[h]
    \centering
    \sidesubfloat[]{\includegraphics[width=0.45\linewidth,frame]{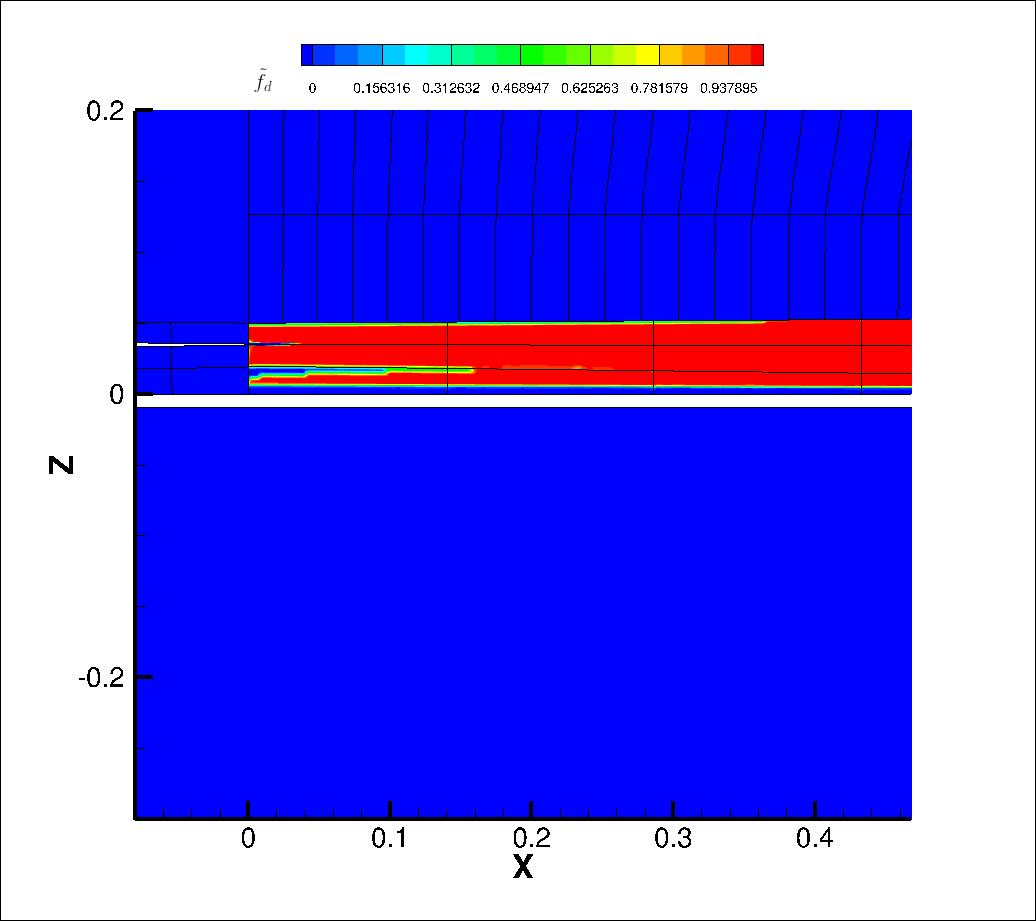}\label{fig:fd:a}}
    \hfill
    \sidesubfloat[]{\includegraphics[width=0.45\textwidth,frame]{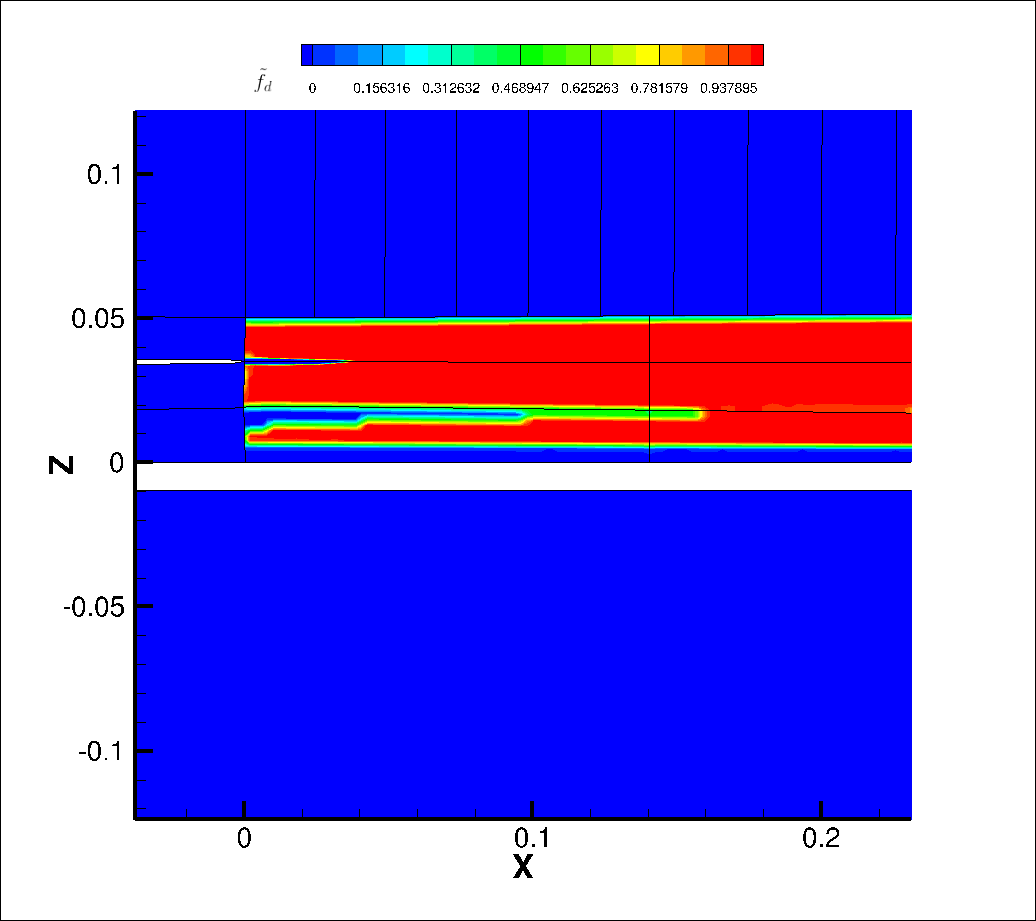}\label{fig:fd:b}}
    \caption{Modified shielding function $\Tilde{f}_d$ in X-LES blocks (a) and close-up on the airfoil trailing edge (b).}
\end{figure}

For the sake of completeness, the contour of the modified shielding function ($\Tilde{f}_d$) is also reported in Figs. \ref{fig:fd:a}-\ref{fig:fd:b}.
The X-LES blocks which enclose the airfoil wake are almost completely simulated in LES mode. On the contrary, the RANS mode is correctly switched on inside the flat plate boundary layer.
From Fig. \ref{fig:fd:b} it is also evident that two short RANS regions are present in close proximity to the airfoil trailing edge and between the airfoil and the flat plate respectively. 
The former is ascribed to the relatively low level of local grid resolution, which is unable to resolve the fine mixing of the flows coming from the upper surface and the lower surfaces of the airfoil.
The latter, instead, is attributed to the lack of relevant turbulent phenomena in the first portion of the X-LES domain located halfway between the two bodies: the flat plate boundary layer is not thick enough to interact with the airfoil wake locally (see Fig. \ref{fig:eddy:b}) so the low levels of modelled kinetic energy coming from the RANS region (i.e., $X<0$) are simply advected downstream until the real wake-boundary-layer interaction takes place.

\begin{figure}[h!]
    \centering
    \sidesubfloat[]{\includegraphics[width=0.45\linewidth,frame]{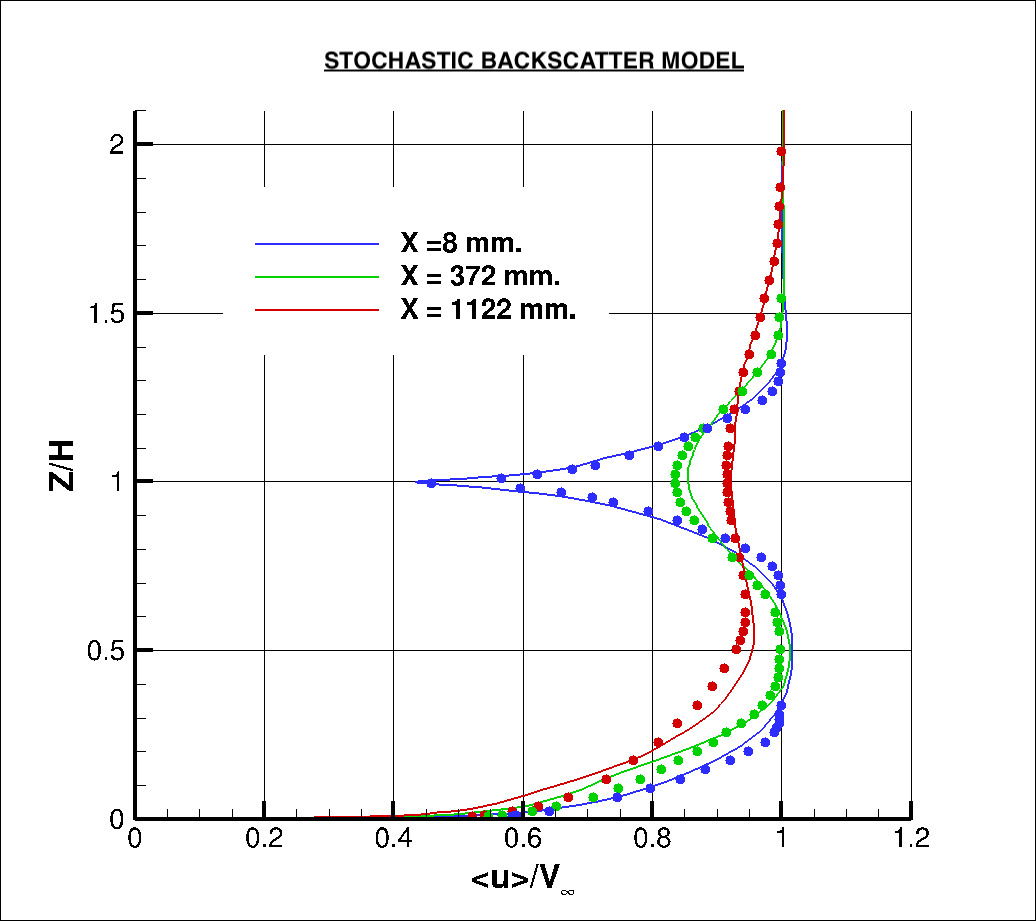}\label{fig:meanvel:a}}
    \hfill
    \sidesubfloat[]{\includegraphics[width=0.45\textwidth,frame]{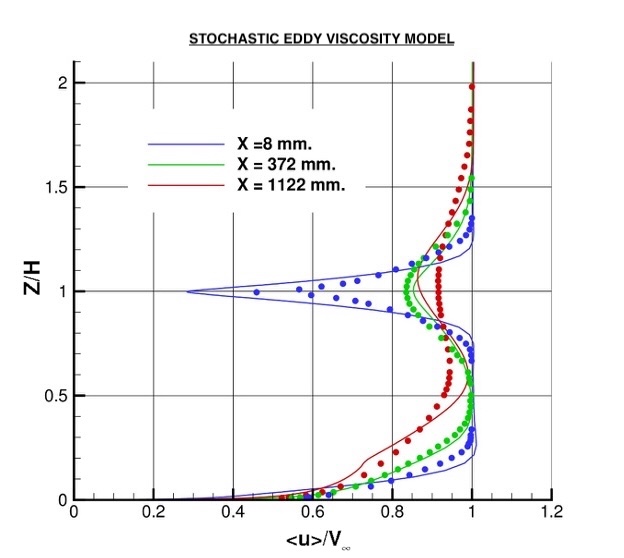}\label{fig:meanvel:b}}
    \caption{Average $X$-velocity profiles at three different sections along the $X$-axis: SBS model (a) vs. SEV model \cite{catalano2019application} (b). Markers denote experimental results \cite{pot1979wake}.}
\end{figure}

The average $X$-velocity profiles at three different sections along the $X$-axis are reported in Fig. \ref{fig:meanvel:a}. For the sake of comparison, the profiles obtained by Catalano \cite{catalano2019application} using the SEV model for the same test case are also shown in Fig. \ref{fig:meanvel:b}. 
It is clear that the overall agreement with the experimental data has substantially improved. In particular, the inflection point in the average velocity profile at $X=\SI{8}{\milli\meter}$ is captured almost perfectly.
A second remarkable difference can be noticed in the last velocity profile ($X=\SI{1122}{\milli\meter}$). Catalano \cite{catalano2019application} found a wavy trend which, in some way, resembles the shape of the average velocity profile at $X=\SI{372}{\milli\meter}$.
This is believed to be an indirect effect of the grey area phenomenon: since resolved turbulence develops with a certain spatial delay downstream of the airfoil trailing edge, there is a region of the flow field where the modelled component of the Reynolds stress tensor is relatively small (due to the switch to LES mode) but where, at the same time, the resolved component of the same tensor is practically zero.
As a consequence, the flow coming from the RANS region of the computational domain proceeds downstream with a much reduced level of diffusion. This delays the mixing between the airfoil wake and the flat plate boundary layer.  

\section{Conclusions}
\label{sec:conclusions}
The SBS model has been incorporated into CIRA's in-house CFD code, UZEN. 
The multi-block nature of the solver has allowed for a zonal implementation of the method: the user can select the blocks which are to be simulated in X-LES mode; the remaining ones are automatically treated in RANS mode.

Particular attention was devoted to the dissipative and dispersive properties of the discretisation scheme for the convective terms in the governing equations.
The approach proposed here consists in the blend between the second-order Low-Dissipation Low-Dispersion (LD2) scheme developed by L\"owe et al. \cite{lowe2016low} and a first-order upwind scheme. The blending parameter is the modified shielding function, which dictates the simulation mode locally employed (i.e., RANS or LES). The standard second-order central scheme with JST-type scalar artificial dissipation is used by default in RANS blocks.

Since the formulation of the shielding function requires the wall distance field, an algorithm for its computation which ensures continuity in all of the blocks in the computational domain has been implemented. 
This is based on the solution of a Poisson equation, which was carried out in a finite-difference framework using the Accelerated Over-Relaxation (AOR) method. 

Since the model constant which appears in the definition of the LES filter width is known to be sensitive to the numerical discretisation employed, the SBS method has been calibrated again on the classic decay of isotropic homogeneous turbulence. The model constant derived from such analysis is only slightly higher than the one computed by Kok \cite{kok2017stochastic} using a fourth-order low-dispersion symmetry-preserving scheme.

Finally, an application of the method to the mixing co-flow between an airfoil wake and a zero-pressure-gradient turbulent boundary layer has been discussed. The focus was on the formation of local instabilities in the free shear layer developing downstream of the airfoil trailing edge. 
The SBS model has proved to be effective in removing the grey area phenomenon, namely the spatial delay in the formation of fluctuating turbulent structures when the flow transitions from a RANS to an LES (or hybrid) region of the computational domain.
The mitigation of the grey area issue has also resulted in a closer agreement of the average $X$-velocity profiles with the experimental data.
In particular, the inflection point in the first velocity profile along the $X$-axis is captured almost perfectly.

\clearpage
\section*{Appendix: variance preservation in the discretised Langevin equations}
The discretised one-dimensional Langevin equation (in primitive form) is reported below for the sake of clarity:
\begin{equation}
    \xi^n + \dfrac{\tau}{2\,\delta t}\left(3\xi^n-4\xi^{n-1}+\xi^{n-2}\right) =F_c\,\sqrt{\dfrac{2\,\tau}{\delta t}} \eta^n
    \label{eq:2ordback}
\end{equation}
\noindent
No flow is initially considered ($\boldsymbol{u}=0$), and the subscripts $ijk$ have been dropped for notational simplicity. Defining the parameter $a\equiv \delta t/\tau $ and rearranging Eq. \eqref{eq:2ordback}, the following expression can easily be obtained:
\begin{equation}
    \left(2a+3\right)\xi^n= 4\,\xi^{n-1} - \xi^{n-2}+F_c\,\sqrt{8a}\,\eta^n
    \label{eq:disc_Lang_simp}
\end{equation}
\noindent
By computing the variance of both sides of the previous equation, one gets:
\begin{multline}
    \left(2a+3\right)^2\mathrm{Var}\{\xi^n\}= 16\,\mathrm{Var}\{\xi^{n-1}\} + \mathrm{Var}\{\xi^{n-2}\}+8a\,F_c^2\,\mathrm{Var}\{\eta^n\} + \\- 8\,\mathrm{Cov}\{\xi^{n-1},\xi^{n-2}\} - 2\sqrt{8a}\,F_c\,\mathrm{Cov}\{\xi^{n-2},\eta^n\} + 8\sqrt{8a}\, F_c\, \mathrm{Cov}\{\xi^{n-1},\eta^n\}
    \label{eq:proof}
\end{multline}
\noindent
In accordance with the theoretical discussion in Section \ref{subsec:SBS}, the following relations must hold:
\begin{equation}
    \mathrm{Var}\{\xi^n\} = \mathrm{Var}\{\xi^{n-1}\}=\mathrm{Var}\{\xi^{n-2}\} = \mathrm{Var}\{\eta^n\}=1
    \label{eq:identity1}
\end{equation}
\begin{equation}
    \mathrm{Cov}\{\xi^{n-1}\eta^n\} = \mathrm{Cov}\{\xi^{n-2}\eta^n\} = 0
    \label{eq:identity2}
\end{equation}
\noindent
Since $\eta$ is independently generated at every physical time step, its current value is supposed not to be correlated with the values of $\xi$ at previous time steps.
By defining $X$ such that $X\equiv\mathrm{Cov}\{\xi^{n-1},\xi^{n-2}\}$ and by substituting the identities reported in Eqs. \eqref{eq:identity1}-\eqref{eq:identity2}, Eq. \eqref{eq:proof} becomes:
\begin{equation}
    \left(2a+3\right)^2= 16 + 1+8a\,F_c^2 - 8X
    \label{eq:firsteqLang}
\end{equation}
\noindent
The above equation must be solved for $F_c$ and $X$. Another relation is thus needed to obtain a problem in closed form. To this aim, Eq. \eqref{eq:disc_Lang_simp} is rearranged as:
\begin{equation}
    (2a+3)\,\xi^n -4\,\xi^{n-1} = -\,\xi^{n-2}+F_c\sqrt{8a}\,\eta^n
\end{equation}
\noindent
Again, the variance of both sides can be computed:
\begin{equation}
    (2a+3)^2+16-8(2a+3)\,\mathrm{Cov}\{\xi^n,\xi^{n-1}\} = 1 + 8a\,F_c^2
\end{equation}
\noindent
Obviously, $\mathrm{Cov}\{\xi^n,\xi^{n-1}\} =\mathrm{Cov}\{\xi^{n-1},\xi^{n-2}\} = X$ since the time step is fixed. A system of two equations in two unknowns is finally obtained:
\begin{equation}
    \begin{cases}
        8a\,F_c^2-\quad\;\;8X &= (2a+3)^2-17\\[3pt]
        8a\,F_c^2+\;8(2a+3)X &= (2a+3)^2+15
    \end{cases}
\end{equation}
\noindent
It is now trivial to derive an expression for the correction factor $F_c$:
\begin{equation}
    F_c = \sqrt{\dfrac{(1+a)(4+a)}{2(2+a)}}
\end{equation}
If the convective term is included in Eq. \eqref{eq:2ordback}, variance preservation cannot be proven exactly \cite{kok2017stochastic}. However, global conservation up to $\mathcal{O}(\delta t)^3$ can be demonstrated if a skew-symmetric discretisation is employed. The interested reader is referred to \cite{kok2017stochastic} for further details. 

\section*{Acknowledgments}
A. Passariello thanks Dr. J. C. Kok (Netherlands Aerospace Centre, NLR) for his unwavering patience and availability in clarifying the many doubts that arose during the course of the work leading to this paper. The authors acknowledge the CINECA award under the ISCRA initiative, for the availability of high performance computing resources and support.

\bibliography{sample}

\end{document}